\documentclass[prb,twocolumn,floatfix,showpacs,superscriptaddress,linenumbers]{revtex4}
\usepackage{graphics}
\usepackage{epsfig}
\usepackage{color}
\usepackage{amsfonts}
\usepackage{amsmath}
\usepackage{amssymb}
\usepackage{amsthm}
\usepackage{hyperref}
\usepackage{wasysym}

\begin{document}

\newcommand{\noteKP}[1]{{\color{magenta}\bf [KP: #1]}}
\newcommand{\noteFP}[1]{{\color{cyan}\bf [FP: #1]}}
\newcommand{\noteCH}[1]{{\color{green}\bf [CH: #1]}}

\title{Interplay of charge and spin fluctuations of strongly interacting electrons on the kagome lattice}

\begin{abstract}
We study electrons hopping on a kagome lattice at third filling described by an extended Hubbard Hamiltonian with on-site and nearest-neighbour repulsions in the strongly correlated limit. 
As a consequence of the commensurate filling and the large interactions, each triangle has precisely two electrons in the effective low energy description, and these electrons form chains of different lengths. 
The effective Hamiltonian includes the ring exchange around the hexagons as well as the nearest-neighbor Heisenberg interaction. 
Using large scale exact diagonalization, we find that the effective model exhibits two different phases: If the charge fluctuations are small, the magnetic fluctuations confine the charges to short loops around hexagons, yielding a gapped charge ordered phase. 
When the charge fluctuations dominate, the system undergoes a quantum phase transition to a resonating plaquette phase with ordered spins and gapless spin excitations. 
We find that a peculiar conservation law is fulfilled: the electron in the chains can be divided into two sublattices, and this division is conserved by the ring exchange term.
\end{abstract}

\author{Frank Pollmann}
\affiliation{Max-Planck-Institut f\"ur Physik komplexer Systeme, 01187 Dresden, Germany}	
\author{Krishanu Roychowdhury}
\affiliation{Max-Planck-Institut f\"ur Physik komplexer Systeme, 01187 Dresden, Germany}	
\author{Chisa Hotta}
\address{Kyoto Sangyo University, Department of Physics, Faculty of Science, Kyoto 603-8555, Japan}
\author{Karlo Penc}
\affiliation{Institute for Solid State Physics and Optics, Wigner Research
Centre for Physics, Hungarian Academy of Sciences, H-1525 Budapest, P.O.B. 49, Hungary}

%\
%\pacs{
%71.10.Fd, 	%Lattice fermion models (Hubbard model, etc.)
%75.10.Jm, 	%Quantized spin models
%75.50.Dd 	%Nonmetallic ferromagnetic materials
%}
\maketitle
\date{\today}
\maketitle
\narrowtext

\section{Introduction}\label{intro}
Strongly correlated systems on frustrated lattices can exhibit very interesting physics. 
The competition between different interactions often results in multiple low-energy states which are degenerate or nearly degenerate with each other. 
Consequently, quantum fluctuations  become very important at low temperatures and can lead to emergent phases of matter with exotic properties. 
\par
One such example is the pyrochlore lattice, which hosts collective excitations that form magnetic monopoles.\cite{moessner} 
At very low temperatures, where quantum fluctuations become important, these models have shown to stabilize an artificial quantum electrodynamics, 
supporting magnetic and electric charges as well as linearly dispersing, gapless excitations (photons).\cite{moessner2003,hermele2004,sikora2012,Benton2012}
Another exciting example is the quantum Heisenberg model on the kagome lattice 
\cite{Mendels2011}, 
which is believed to exhibit a $\mathbb{Z}_2$ topological liquid ground state, 
carrying anyonic excitations.\cite{white, depenbrock2012, balents2012} 
Recently, indications for the existence of a $\mathbb{Z}_3$ spin liquid have been observed when a finite magnetic field is applied.\cite{Nishimoto2013} 
Both examples are systems involving localized spins. 
The effect of frustration on charge degrees of freedom has received less attention so far. 
Still, in a number of recent works, it has been shown that models at partial filling (i.e., itinerant systems) on frustrated lattices support fractional charges in two- and three dimensions. \cite{fulde2001, runge2004, pollmann2006, obrien2010} 
This fact is quite interesting as there exist only few examples of models that support fractional charges in higher dimensions. 
One very interesting and important question is about the interplay of spin- and charge degrees of freedom.
For a checkerboard lattice model at quarter filling, the interplay between charge and spin degrees of freedom can stabilize a robust insulating resonating singlet-pair crystal phase.\cite{Poilblanc2007,Trousselet2008}
Previous studies on the kagome at different filling factors have revealed rich phase diagrams including various symmetry broken as well as topological phases.   \cite{wen2010,Indergand2006} 
At filling factor one sixth, a new mechanism for ferromagnetism on the kagome lattice was found.\cite{pollmann2008}

\par
In this paper, we study the interplay between spin- and charge degrees of freedom on the kagome lattice at a filling factor of $n=2/3$. 
Most interestingly, we find that the spin-fluctuation in the model can drive the systems through a phase transition into a charge ordered phase. 
We start from an extended Hubbard model on the kagome lattice for which we derive a low energy Hamiltonian using degenerate perturbation theory. 
By considering different limiting cases, we obtain some insight into the different phases of the model. 
In the limit where antiferromagnetic spin fluctuations dominate, a ``short loop'' phase is formed in which the charges align around hexagons. 
On the other hand, if the charge fluctuations dominate, we find a ``plaquette ordered'' ground state. 
For the latter limit (no spin-fluctuations),  we find a very peculiar conservation law, namely, the Hamiltonian conserves the magnetization on dynamic sublattices. 
To get a picture of the whole phase diagram, we perform a large scale exact diagonalization study of cluster up to $N=36$ sites in which we calculate the energy spectrum and different correlation functions from which we conclude about the phase diagram. 
This paper is organized  as follows:
In Sec.~\ref{sectwo}, we introduce the model Hamiltonian and derive the low-energy effective Hamiltonian. 
We then consider the limiting cases  in Sec.~\ref{secthree}, allowing us to make some statements about some corners of the phase diagram. 
The full phase diagram is then evaluated using exact diagonalization in Sec.~\ref{secfour} 
We conclude by giving a short summary and outlook in Sec.~\ref{sum}.

\section{Model Hamiltonian}\label{sectwo}
Here we consider the extended Hubbard model on the kagome lattice
with on-site and nearest-neighbor repulsive interactions, $U$ and $V$, respectively, 
with a Hamiltonian
\begin{equation}
{\mathcal H}=-t\sum_{\langle
i,j\rangle,\sigma}\left(c_{i\sigma}^{\dag}c^{\vphantom{\dag}}_{j\sigma}
+ \text{H.c.}\right) \\
  +V\sum_{\langle
i,j\rangle}n_{i}n_{j}+U\sum_{i}n_{i\uparrow}n_{i\downarrow}.
\label{eq:extended_hub}
\end{equation}
The operators $c^{\vphantom{\dag}}_{i\sigma}$ ($c_{i\sigma}^{\dag}$) 
annihilate (create) an electron with spin $\sigma$ on site $i$, 
$n_{i}=n_{i\uparrow}+n_{i\downarrow}$ is the electron number operator 
with $n_{i\sigma}=c_{i\sigma}^{\dag}c^{\vphantom{\dag}}_{i\sigma}$, 
and the notation $\langle i,j\rangle$ refers to pairs of nearest neighbors. 
Throughout this paper, we focus on the case of one-third filling, in the strongly correlated regime, 
where $|t| \ll V < U$. For this filling there are two electrons on each triangle on average (i.e., the total number of electrons is $N_e=2N/3$, where $N$ is the number of lattice sites, providing a filing factor $n=N_e/N=2/3$).

\begin{figure}
\begin{centering}
\begin{tabular}{ccc}
\includegraphics[width=80mm]{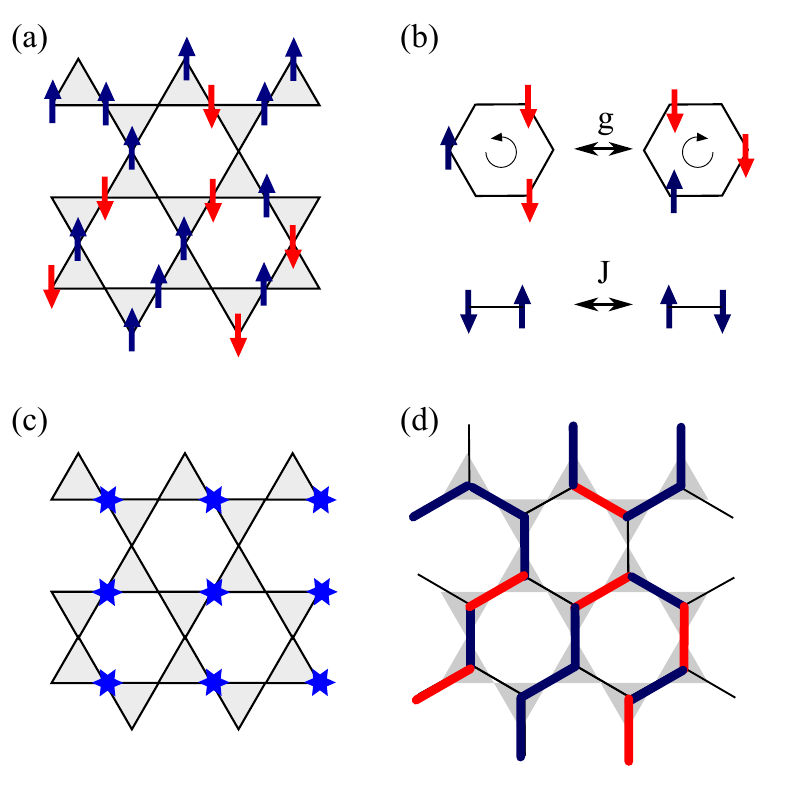}&
\end{tabular}\par
\end{centering}
\caption{(a) A configuration satisfying the constraint of zero or one electron per site and two electrons of arbitrary spin per triangle. (b) Quantum fluctuations allow tunneling between different degenerate configurations of spins and charge: Ring exchange with amplitude $g$ and spin-exchange with an amplitude of $J$. The ring exchange processes change the number of electrons on the starred (blue online) sublattice shown in panel (c) always by two. This is crucial for gauging away the sign of $g$ explained in the text. (d) All allowed configurations can be expressed in terms of a two-colored fully packed loop model on the honeycomb lattice; as an example, we show a representation of the configuration presented in (a). 
\label{dimerpanel}}
\end{figure}

In the strong coupling limit, when the hopping amplitudes are set to zero and $0<V<U$, the energy is minimized if there are {\it exactly} two electrons per each triangle of the kagome lattice with no double occupancy -- this is analogous to the case of magnetite as discussed in Ref.~ \onlinecite{Anderson1956}. 
An example configuration fulfilling these constraints is shown in Fig.~\ref{dimerpanel}(a). 
The number of such configurations is macroscopically degenerate: in addition to the trivial $2^{2N/3}$ spin degeneracy, the number of charge configurations also grows exponentially with the system size.
The ground state configurations on the kagome lattice can be mapped  to two-colored fully packed loop configurations on the honeycomb lattice [particles are sitting here on links of the honeycomb lattice, and the two different colors encode the spin orientation, see Fig.~\ref{dimerpanel}(d)]. 
The charge degrees of freedom (i.e, neglecting the color)  can be equivalently described by a dimer model of the honeycomb lattice by simply replacing empty bonds by occupied ones and occupied by empty ones.
Using this mapping, the degeneracy of different charge configurations can be calculated exactly using Pfaffians,\cite{fisher} and is given as $\sim 1.1137^N$ for a honeycomb lattice with $N$ bonds (corresponding to $N$ kagome sites).
%
%At the same time, a Pauling-like\cite{Pauling} estimate gives  $\propto (4/3)^{(N/3)}\approx 1.1006^N$, slightly underestimating the exact value.
%
The total degeneracy is then the product of the spin degeneracy and the charge degeneracy, i.e., $\sim 2^{2N/3} \times 1.1137^N$.
\par
The macroscopic ground-state degeneracy is lifted when quantum fluctuations are taken into account.
The effective Hamiltonian that connects a manifold of degenerate states can be obtained from a perturbative expansion of the Hamiltonian Eq.~(\ref{eq:extended_hub}) in $t/V$ and $t/(U-V)$. 
By keeping only the lowest order of non-vanishing terms, one obtains the effective Hamiltonian as a sum of two parts:
\begin{equation}
{\mathcal H}_{\text{eff}}= {\mathcal H}_{\text{ring}}+ {\mathcal H}_{\text{spin}} .
\label{eq:Hlow}
\end{equation}
The first term describes a ring exchange of three electrons occupying every other site on a hexagon of the kagome lattice, and is given as
\begin{equation}
{\mathcal H}_{\text{ring}}=
-g
\sum_{\left\{\hexagon\right\}} h_{\hexagon}
\label{eq:Hhex}
 \end{equation}
with an effective ring-exchange amplitude $g=6t^{3}/V^{2}$ and
\begin{equation}
h_{\hexagon} =\sum_{\left\{ \sigma, \sigma',\sigma''\right\}}
 \left( c^{\dagger}_{n\sigma''}c^{\vphantom{\dagger}}_{m\sigma''}c^{\dagger}_{l\sigma'}c^{\vphantom{\dagger}}_{k\sigma'}c^{\dagger}_{j\sigma}c^{\vphantom{\dagger}}_{i\sigma} + \text{H.c.}\right).\label{eq:ring}
 \end{equation}
The sum is performed the over sites of the hexagon and all spin orientations. 
The indices $i,j,k,l,m,n$ are sites oriented clockwise on a hexagon yielding the dynamics sketched in Fig.~\ref{dimerpanel}(b), i.e., three electrons hop collectively either clockwise or counter-clockwise around the hexagons. 
Clearly, this ring exchange process preserves the number of electrons on each triangle, and if applied to a state that belongs to the ground state manifold, the resulting state will also belong to the same manifold. 
Notice that the fermionic sign in expression for ${\mathcal H}_{\text{ring}}$ can be gauged away, yielding a bosonic model. \cite{obrien2010}
Furthermore, the overall sign of $g$ can be transformed by a simple gauge transformation which multiplies all configurations with the factor $i^{N_{\text{star}}}$, where $N_{\text{star}}$ is the number of electrons on the sublattice shown in Fig.~\ref{dimerpanel}(c). 
\par
The second term in the effective Hamiltonian (\ref{eq:Hlow}) is the nearest neighbor Heisenberg exchange 
\begin{equation}
{\mathcal H}_{\text{spin}} = J\sum_{\langle i,j\rangle}\left(2S_{i}S_{j}-\frac{1}{2}n_{i}n_{j}\right),
\label{eq:Hspin}
\end{equation}
where 
\begin{equation}
 J=\frac{2t^2}{U-V}+\frac{2t^3}{V^2} \;.
 \label{eq:Jexch}
\end{equation}
In the ground state manifold, each electron has two occupied neighboring sites (one on each of the two corner sharing triangles), so that closed loops are formed in a system with periodic boundary conditions. 
These loops are like spin chains and the exchange Hamiltonian ${\mathcal H}_{\text{spin}}$  acts on the spins of the electrons in these closed loops without modifying the charge configuration. 
The length of the loops is always even, and the shortest looplength is six. 
For $U\gg V$ the first term in Eq.~(\ref{eq:Jexch}), proportional to $t^2$, becomes small compared to the term that is $\propto t^3$, so that the sign of the exchange depends on the sign of the hopping amplitude $t$, allowing antiferromagnetic as well as ferromagnetic exchanges (analogously to the one-dimensional case considered in Ref.~\onlinecite{penc96}). %
One important aspect of effective model, Eq.~(\ref{eq:Hlow}), is that $g$ and $J$ can be regarded as nearly independent variables: One can tune the value of $J$ by changing $U$, without affecting $g$. 
We can reparametrize them by a single variable 
\begin{equation}
  \alpha = \frac{|g|}{|g|+|J|}\;,
\label{eq:alpha}
\end{equation}
which falls within $0\leq\alpha\leq1$. 
In the limiting case $\alpha\rightarrow 0$ we can neglect the effect of the ring exchange term.
This happens when the $V\rightarrow U$ (but still $V<U$), as due to the divergence in Eq.~(\ref{eq:Jexch}), the effective exchange becomes much larger than the $g$. In the limit of $V\ll U$, the $\alpha$ approaches   $\alpha \rightarrow 3/4$ as 
\begin{equation}
 \alpha = \frac{3}{4} - \frac{3 V^2}{16 t U} + O(1/U^2) \;.
\label{eq:alpha2}
\end{equation} 
However, the overall behavior is not so simple and in the following we consider the different signs of the hopping $t$ separetely: Figures 2(a) and (b) show the contour plot of $\alpha$ on the plane $U$ and $V$ (more precisely, $t/U$ and $t/V$) for $t>0$ and $t<0$, respectively. 

For $t>0$, the exchange is always antiferromagnetic and $0<\alpha\leq3/4$. 
To increase the $\alpha$ from $\alpha=0$ limit at $U=V$, we need to increase $U$ compared to $V$, and from Eq.(\ref{eq:alpha2}) 
we find the upper bound as 3/4.
 
The situation is more involved when $t<0$: 
we encounter both ferromagnetic and antiferromagnetic $J$, and the $\alpha$ can take the values $0<\alpha \leq 1$. 
The line $\alpha =1$ in Fig.~2(b) is determined by the antiferromagnetic $\propto t^2$ term, canceling the ferromagnetic $\propto t^3$ term in $J$ [see Eq.~(\ref{eq:Jexch})]. 
When the effective exchange is ferromagnetic, the values of $\alpha$ are limited to $3/4\leq\alpha<1$: 
as $U$ increases from the $\alpha=1$-line, $\alpha$ decreases down to 3/4 in the $U \gg V$ (Eq.(\ref{eq:alpha2})). 
The antiferromagnetic exchange is realized by decreasing $U$ from the $\alpha=1$ line toward $U=V$, which decreases $\alpha$ down to zero. 
These factors indicate that, by a suitable choice of the values of the interactions and hoppings, we can select antiferromagnetic exchange with arbitrary value of $0<\alpha<1$. 

Here, let us mention that we can link our model to the "flat band" ferromagnetism: the Hubbard model with $t>0$ and $n\leq 1/3$ has been proven to be ferromagnetic for any $U>0$ and $V=0$,\cite{Mielke1992,MielkeTasaki1993}. Furthermore, the $n=1/3$ case with $|t|\ll V\ll U$ is also proven to exhibit a ferromagnetic ground state\cite{pollmann2008} -- thus here we extend the possibility of a ferromagnetic ground state also to filling factor $n=2/3$. 

In the remainder of the paper, we will consider how the effective Hamiltonian lifts the degeneracy of the ground state manifold 
in the strong coupling limit. 

\begin{figure}
\begin{centering}
\includegraphics[width=8.5truecm]{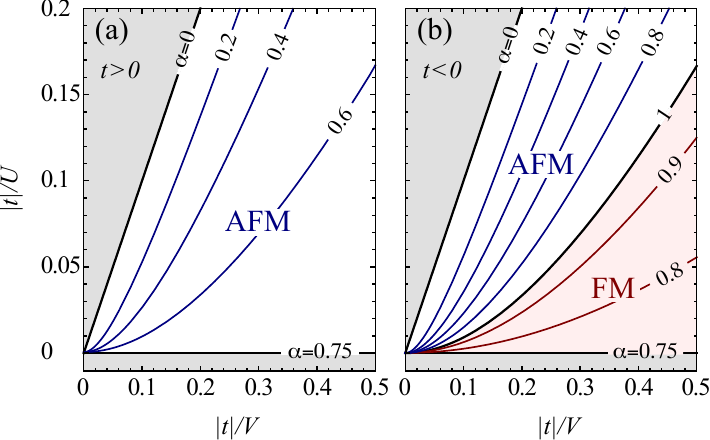}
\end{centering}
\caption{\label{fig:alphavsUV}
The values of the parameter $\alpha$ defined in Eq.(\ref{eq:alpha}) 
as contourplots on the plane of $|t|/U$ and $|t|/V$. 
(a) For $t>0$ the exchange is always antiferromagnetic, (b) while for $t<0$ the exchange becomes ferromagnetic (red region) for large on-site repulsion $U$, as it follows from the perturbative expansion up to third order in the hopping, Eq.~(\ref{eq:Jexch}). 
The effective Hamiltonian describes the region close to the origin in the unshaded region $U>V$ and $U>0$.}
\end{figure}

\section{Limiting cases}\label{secthree} 

It is instructive to first consider the two terms in the effective Hamiltonian Eq.~(\ref{eq:Hlow}) separately. 
This corresponds to the setting $\alpha=1$ and $\alpha=0$. 
Understanding these two limits will help us get a picture of the full phase diagram.
\subsection{ Plaquette phase of the ring exchange Hamiltonian ($\alpha=1$)}
At $\alpha=1$, the effective Hamiltonian reduces to ${\mathcal H}_{\rm ring}$ given by Eq.~(\ref{eq:Hhex}). 
We first use the mapping to the two-color fully packed loop model [see Fig.~\ref{dimerpanel}(d)] to understand the charge dynamics.
Second, we discuss a hidden symmetry of this model that yields large degeneracies.  
\subsubsection{Resonating plaquettes and winding numbers}

When the spins of the electrons are all pointing in the same direction (e.g. up, $S=S^z=S^{\text{max}}_{\text{tot}}$), the spins can be omitted and the relevant degrees of freedom are the positions of the charges. 
As described in Sec.~\ref{sectwo}, the charge problem can be mapped to a  dimer model on the honeycomb lattice.
The quantum-dimer model on the honeycomb lattice with resonances on the neighboring disjunct hexagons has been shown to have a gapped, plaquette ordered  ground state with an off--diagonal order parameter \cite{Moessner01c} -- the so called ``Plaquette Phase''.  
The ground state is three-fold degenerate in the thermodynamic limit, breaking the translational symmetry of the lattice.
The plaquette phase, when mapped back to the kagome lattice model, hosts electrons resonating around the hexagons  (see right panel in Fig.~\ref{gspt}).
Then, one can define conserved quantities (winding numbers) that can be used to classify the states in the Hilbert space, similar to the quantum-dimer model on a square lattice.\cite{Rokhsar88} 
In our case, ${\mathcal H}_{\rm ring}$ conserves the number of electrons along the straight lines parallel to the edges of the hexagons in the kagome lattice. 
Actually, the number of linearly independent winding numbers is only two when the system is put onto a torus (i.e., when considering standard periodic boundary conditions).
The Hilbert space is divided into subspaces (sectors), as only  states having the same winding numbers are connected by ${\mathcal H}_{\rm ring}$.
\subsubsection{Hidden conservation law}
%*%*%*%*%*%*%*%*%*%*%*%*%*%*
\begin{figure}
\begin{centering}
\includegraphics[width=80mm]{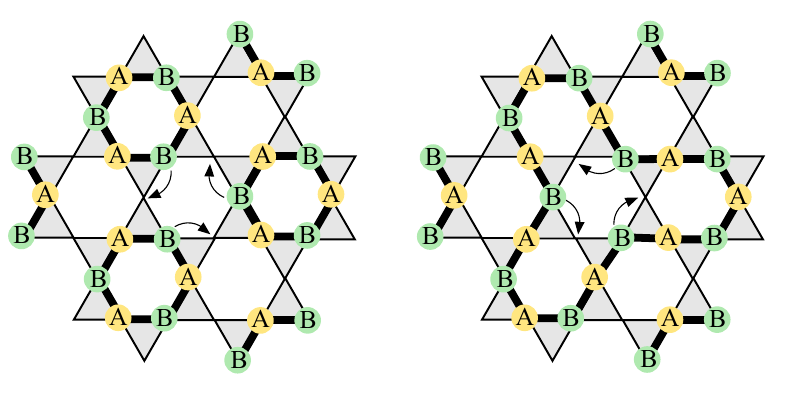}
\end{centering}
\caption{Dynamic sublattice structure defined on a loop of charges, consisting of even number of sites. 
The effective ring exchange denoted by arrows 
preserves the magnetization on each of the dynamic sublattices $A$ and $B$ (see Appendix \ref{sec:dynsublatt} for details). }
\label{sublattices}
\end{figure}
%*%*%*%*%*%*%*%*%*%*%*%*%*%*

In what follows, we investigate the effect of the ring exchange on the spins of electrons in the loops. 
We will show that, in addition to the trivial conservation of total $z$-component $S^{z}_{\text{tot}}$, and the total spin $S_{\text{tot}}$, a {\it hidden conservation law} emerges which we describe below. 

Let us introduce the {\it ``dynamic two-sublattices''} which are defined on top of the electron loops (Fig.~\ref{sublattices}): 
starting from an arbitrary choice of loop configuration of electrons, we assign the bipartite labels, $A$ and $B$, to each electron in the loop -- this is possible because loops consist of even number sites.
To be more precise, the rules to construct the dynamic sublattice are simple: (i) neighboring electrons have different sublattice labels and (ii) next-nearest neighbor electrons have the same sublattice label  on each hexagon of the kagome lattice. 
For example, we find hexagon configurations which are completely filled ($ABABAB$ as we go around the hexagon), or depleted hexagons as $AOAOAO$, $ABAOAO$, and $ABABAO$ (here the $O$ denotes empty sites). 
Once we have the configuration which fulfills (i) and (ii), the effective Hamiltonian ${\mathcal H}_{\rm ring}$ conserves this rule: After operating ${\mathcal H}_{\rm ring}$ arbitrary times, one finds that the bipartite configuration is perfectly kept -- each electron can be assigned not only the spin, but also the label denoting the dynamic sublattice. The proof of this conjecture is given in the Appendix \ref{sec:dynsublatt}.
 Here, notice that even if we come back to the same charge configuration at some point, the sites that were occupied by $A$ electrons could all be replaced to $B$ and vise versa. 
In fact, even though the bipartite sublattice rule is kept in the loops, the absolute location of $A$ and $B$ are not fixed, which is the reason why we call them ``dynamic''. 
\par
Next, we assign the spins to the electrons in the loops. 
Since the electrons in the $A$ and $B$ sublattices never exchange with each other, the total $S^{z}_{A}$ and $S^{z}_{B}$ on each of the dynamic sublattices is a conserved quantity, so it is a good quantum  number. 
This is not only true for the $z$ component, but also for the $\mathbf{S}_A$ and $\mathbf{S}_B$. 
As a consequence, the Hilbert subspaces of a fixed winding number are further divided into sectors that do not mix the spin $S^z$ on the two sublattices. 
For a system with the magnetization $S_{\text max}-n$ we find $\lfloor n/2+1\rfloor$ disconnected  sectors, e.g., if we take $S_{\text max}-2$, namely flipping two spins from a fully spin polarized configuration, we can either flip both spins on one of the dynamic sublattice or one on each yielding two disjoint sectors. 
In fact, by keeping $S^{z}_{\text{tot}}=S^{z}_{A}+S^{z}_{B}$, the number $\lfloor n/2+1\rfloor$ is equal to the number of ways one can add two integers. 
\par
The hidden conservation law yields a spin degeneracy. 
To show this, we introduce the $P_{AB}$ operator that exchanges the $A$ and $B$ sublattice labels of the electron operators (note that this operator does not change the charge configuration). 
Since $P_{AB}^2=1$, the wave functions are either even or odd with respect to $P_{AB}$, with eigenvalues $\pm 1$. 
The ring exchange Hamiltonian and the $\mathbf{S}_A+\mathbf{S}_B$ commute with the $P_{AB}$, while the $\mathbf{S}_A-\mathbf{S}_B$ does not. 
However, we can define the operator, 
\begin{equation}
Q^{\mu\nu} = (S_A^\mu-S_B^\mu)(S_A^\nu-S_B^\nu) 
\end{equation}
which commutes with both the $P_{AB}$ and the ${\mathcal H}_{\text{ring}}$ (where $\mu,\nu=x,y,z$). The $Q^{\mu\nu}$ has nonvanishing matrix elements between total spin states that differ by 2, so these states are also degenerate in energy (more precisely, the $Q^{\mu\nu}+Q^{\nu\mu}- (2/3) \delta_{\mu\nu} Q^{\eta\eta}$  is a rank two tensor operator). 
For example, applying the $Q^{--}$ to the highest weight state of the maximal spin we create a state that is a linear superposition of the $S_{\text{tot}}^{\text{max}}$ and $S_{\text{tot}}^{\text{max}}-2$, and is degenerate with the $S^{\text{max}}$. 
\begin{figure}
 \begin{centering}
 \includegraphics[width=8.5truecm]{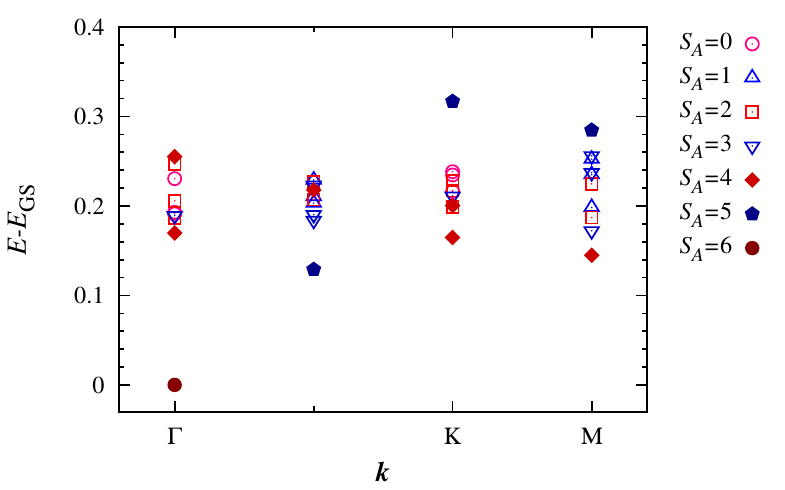}
 \end{centering}
\caption{\label{fig:SA6SB}
Energy spectrum of the $N=36$ cluster at $\alpha=1$. 
We set one of the two dynamic sublattices to be fully polarized as $S_B=6$, 
while varying the spin $S_A$ of the other sublattice. 
The ground state is realized at $\Gamma$-point for $S_A=6$, indicating that both dynamic sublattices hold maximal spins. }
\end{figure}
Regarding the ground state, we find that the total spin on $A$ and $B$ dynamic sublattices is maximal, $S_{A}=S_{B}=N/6$. 
This could be observed explicitly for the $N=36$ site cluster shown in Fig.~\ref{fig:SA6SB}: 
There, we keep one of the two sublattices polarized as $S_B=6$, and vary the total spin of the other sublattice ($S_A$). 
We find that the ground state does in fact have a maximal $S_A$. 
The two ``giant'' spins can be combined to make $S_{\text{tot}}^{\text{max}}-2 m$ spin states that are even with respect to $P_{AB}$, all having the same minimal energy (here $m$ is an integer). 
Similarly, the  $S_{\text{tot}}^{\text{max}}-2 m-1$ are also degenerate, and are odd eigenstates of $P_{AB}$. In other words, in the ground state the two giant spins on the two dynamic sublattices behave as noninteracting spins (except for the parity effect with respect to $P_{AB}$ that disappears in the thermodynamic limit). 
To this end, a qualitative difference between one--third and one--sixth filled case becomes clear. 
In the one--sixth filled case ($n=1/3$), the effective Hamiltonian  $H_{\text{ring}}$ connects all spin configurations within the zero winding sector yielding a ferromagnetic ground state.\cite{pollmann2008}\footnote{This can be traced back to the number of fermions taking place in the effective ring exchange: for odd number of fermions high spin is favoured, while for even number of electrons a low spin state (singlet) is lower in energy, as e.g. in Ref.~\onlinecite{Poilblanc2007} for a quarter filled checkerboard lattice.}

This is no longer the case for the one--third filled case: the ground state is degenerate, and the ferromagnetic state as well as the singlet state are among the ground states. 
However, the dynamic subsystems $A$ and $B$ are still ferromagnetic: our system can be thought of putting together two one--sixth filled systems, each of them living on the $A$ and the $B$ dynamic sublattices of the ground state manifold.

\subsection{Short loop phase ($\alpha=0$)}
\begin{figure}
\begin{centering}
\includegraphics[width=80mm]{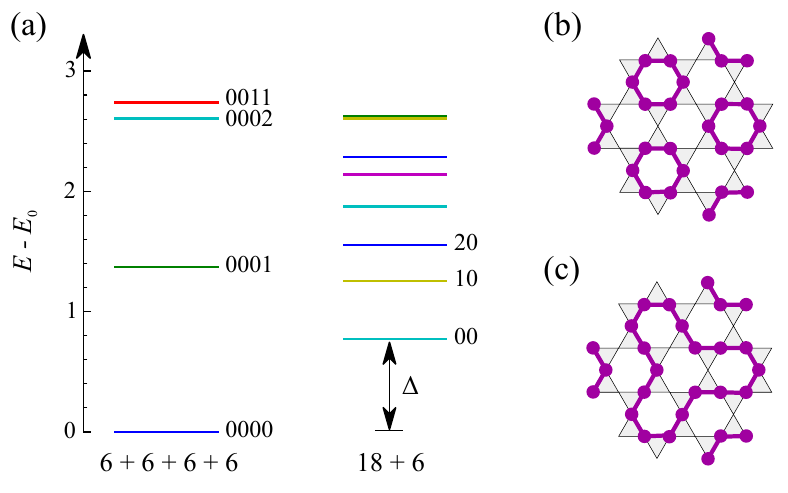}
\end{centering}
\caption{(a) Excitation energies of the spin Hamiltonian ($\mathcal H_{\rm spin}$) 
for two different loop configurations of the $36$-site cluster: 
one consisting of four short-loops of length-six, 
and the other consisting of one 18-site loop and one six-site loop. The numbers next to the levels indicate in which state each loop is, e.g., $0001$ means that three loops are in a ground state and one of them is in the first excited spin state.}
\label{energies}
\end{figure}
At $\alpha=0$, the effective Hamiltonian is reduced solely to ${\mathcal H}_{\text{spin}}$. 
Since the fluctuations of electron occupation vanish, the ground state is the one which minimizes the spin exchange interactions along the closed electron loops.
This is achieved with short hexagonal loops giving rise to a ``Short loop Phase'', shown in the left panel of Fig.~\ref{gspt}. 
The finite size correction for the ground state energy per site of a periodic antiferromagnetic Heisenberg chain of length $L$ is\cite{Woynarovich1987}
\begin{equation}
e_L-e_\infty = -\frac{\pi^2}{12L^2}
\end{equation}
%$e_L-e_\infty\approx-(2\pi)^2/48L^2$,
in a leading order in $1/L$, 
where $e_\infty=1/4-\ln2 \approx -0.4431 $ stands for the energy density in thermodynamic limit. 
Thus, the shorter the loop is, the lower the energy density becomes. The shortest loop on the kagome lattice is of length $L=6$ with energy $e_6 \sim-0.4343$, and these loops arranged in a regular pattern, as shown in Fig.~\ref{energies}(b), with a hexagonal unit cell consisting of 9 sites.  
This ground state is three fold degenerate and breaks the translational symmetry. 
The lowest energy excitation of this charge ordered phase is realized by the formation of a $L=18$-loop out of three adjacent 
hexagonal loops [Fig.~\ref{energies}(c)], with the energy gap, $\Delta=E_{18}-3E_{6}=0.771$. 
In Fig.~\ref{energies}(a) the energies of different charge (or loop) configurations on a 36-site cluster are shown. 
\section{Numerical results}\label{secfour}
We have already found that the two extreme cases of $\alpha=0$ and $\alpha=1$ show different orderings.
The transition between the two phases can be understood by using the analogy to the quantum-dimer model on the honeycomb lattice.\footnote{Even though the effective Hamiltonian~(\ref{eq:Hlow}) 
is more complex than the quantum-dimer model, they both have 
the same symmetry properties with respect to the charge degrees 
of freedom. The short-loop phase can be mapped to the columnar 
phase and the plaquette phases can be identified in both models.}
As discussed in detail in Ref.~\onlinecite{Moessner01c}, the two different orderings have centers of rotation symmetry that lie in distinct places when forming domains of one phase with the other (in fact, the precise nature of the phase transition might be either a first order one, or two phase transitions with coexisting order parameters, as suggested in Ref.~\onlinecite{Ralko2008} for the quantum dimer model on square lattice).
Thus we expect a phase transition between them when tuning the parameter $\alpha$ from 0 to 1.

In order to pin down the transition and to verify the above mentioned two characteristic phases, we employ numerical exact diagonalization on the effective Hamiltonian $H_{\text{eff}}$ on finite clusters of sites $N=27$ and 36. 
We simulate  $H_{\text{eff}}$ within the Hilbert space spanned by the allowed configurations.
Furthermore, we reduce the Hilbert space size by making use of the spatial symmetries  given in the Appendix \ref{sec:IR}. 
The results are summarized in Figure~\ref{gspt} which shows the phase diagram we obtain from our numerical analysis: We observe  a first order phase transition from a ``short loop'' to a ``plaquette ordered'' phase at  $\alpha\approx0.6$. 
Both phases have a charge gap but only the former one has a spin gap. 
The details of the numerical simulations are described below.

However, before presenting our numerical findings, we shall mention that
the strong coupling limit of the Hubbard model (more precisely $tJ$ model) at the same filling as ours, but without the nearest neighbor $V$ term has been discussed in Ref.~\onlinecite{Indergand2006}. It has been found that the ground state is formed by a resonance of two electrons in the singlet state on disjunct triangles, making a crystal. The ground state is twofold degenerate, depending if the resonances are taking place on the up or down pointing triangles. The crucial difference with respect to our model is that the number of electrons in the triangle is not restricted to be precisely two, but only on average. In fact, the electron number strongly fluctuates in the triangles connecting the resonating pairs which costs energy due to $V$ term and leads to destabilization of the state. The Hubbard model for the one-third filling case has been studied in Ref.~\onlinecite{wen2010} using a Hartree-Fock mean field theory which provided a rich phases diagram: our short loop phase can be recognized as the CDWIII phase in their work, while our plaquette phase is missing as the quantum fluctuations stabilizing the resonance of charges are beyond the reach of the Hartree-Fock approach.

\subsection{Anderson tower}

\begin{figure}
 \begin{centering}
 \includegraphics[width=8.5truecm]{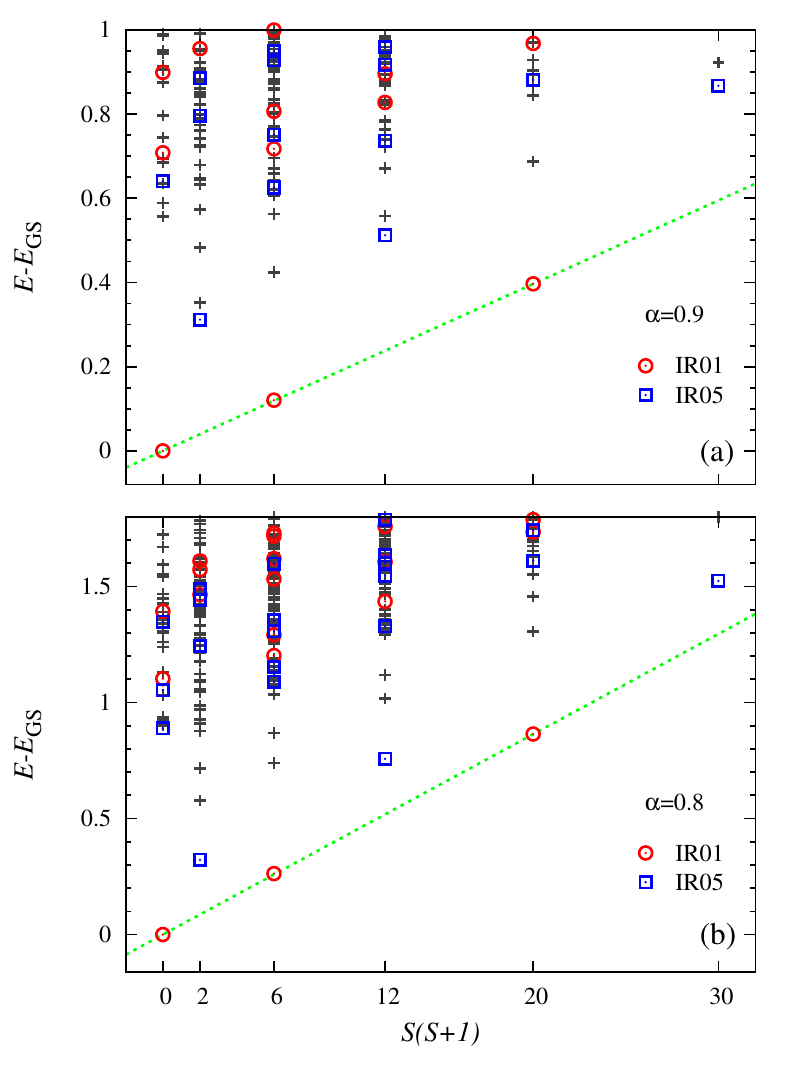}
 \end{centering}
\caption{\label{fig:andtow}
Anderson tower for the $N=36$ cluster for (a) $\alpha=0.9$ and (b) $\alpha=0.8$. 
The energies of the low lying two series of states belonging to different irreducible representations, IR01 and IR05 (see Appendix\ref{sec:IR}), 
which are even and odd with respect to $P_{AB}$, behave proportional to $S(S+1)$. 
As $\alpha$ increases and/or the system size increases, their gradient decreases toward zero, 
at which all these points fall onto the same horizontal line (become degenerate). 
Such behavior is the strong indication of the antiferromagnetic ordering. }
\end{figure}

As for the case of $\alpha=1$, we have seen that the conservation of the spin on the $A$ and $B$ dynamic sublattices leads to a degenerate ground state. 
The spins on the two dynamic sublattices take the maximal value $S_A = S_B = N/6$, and as they do not interact, the states spanned by the two ``giant'' spins constitute the ground state manifold (Fig.~\ref{fig:SA6SB} reveals that states with lower sublattice spin $S_A$ and $S_B$  are higher in energy). 
Once the ${\mathcal H}_{\text{spin}}$ is turned on, the hidden conservation law does not hold any longer, and the spins on the $A$ and $B$ sublattice start to interact with each other: 
\begin{equation}
 \mathcal H_\text{LM}\propto \frac{J}{N} \, {\bf S_{\text{$A$}}} \cdot \bf S_{\text{$B$}} \;,
\end{equation}
similarly to a Lieb-Mattis model, as the kinetic term $g\gg J$ decouples the wave function and each spin on the $A$ sublattice interacts with each spin on the $B$ sublattice with an effective coupling $\propto J/N$. 
Denoting by $S_{\text{tot}}$ the total spin of the system, the energy of this Hamiltonian is simply described as, 
\begin{equation}
 \mathcal E_\text{LM}\propto \frac{J}{N} \left[ S_{\text{tot}}(S_{\text{tot}}+1)-S_A(S_A+1)-S_B(S_B+1)\right] \;. 
\end{equation} 
Indeed, the degeneracies at  $\alpha=1$ are quickly removed when decreasing $\alpha$ (see Fig.~\ref{fig:energy_spectra}). 
When plotted against $S_{\text{tot}}(S_{\text {tot}}+1)$, the spectrum shows low energy states whose energy is $\propto J/N S_{\text{tot}}(S_{\text {tot}}+1)$, as shown in Fig.~\ref{fig:andtow}. 
These states form the Anderson tower, which is the clear signature of an antiferromagnetic ordering.\cite{bernu2}
In this case, the texture of the antiferromagnetic order is quite peculiar, as schematically shown in right part of Fig.~\ref{gspt}: 
a large effective spin-3/2 of resonating charges on a hexagon is surrounded by ``localized'' spin-1/2 electrons. 
The resonating plaquette can occupy either of the three inequivalent hexagon sublattices, 
thus the state is three-fold degenerate regarding the space group symmetries. 
This is reflected in the irreducible content of the states in the Anderson tower in Fig.~\ref{fig:andtow}; 
IR1 and IR5 for the 36 site cluster (see Appendix~\ref{sec:IR} for the whole chart of the irreducible representation). 
With increasing system size, the slope of  the lowest energy levels approaches zero as $1/N$ and become degenerate in the thermodynamic limit. 
The finite size gap of the states above the Anderson tower also goes to zero --- in the case of antiferromagnetic ordering we expect 
that the scaling follows $1/\sqrt{N}$, which, however, could not be checked in our problem due to rapidly growing dimension of the Hilbert space.
The spin excitation spectrum becomes gapless, in contrast to the short loop phase when $\alpha\approx 0$.

\subsection{Energy spectrum}

\begin{figure}
\begin{centering}
\includegraphics[width=8.5truecm]{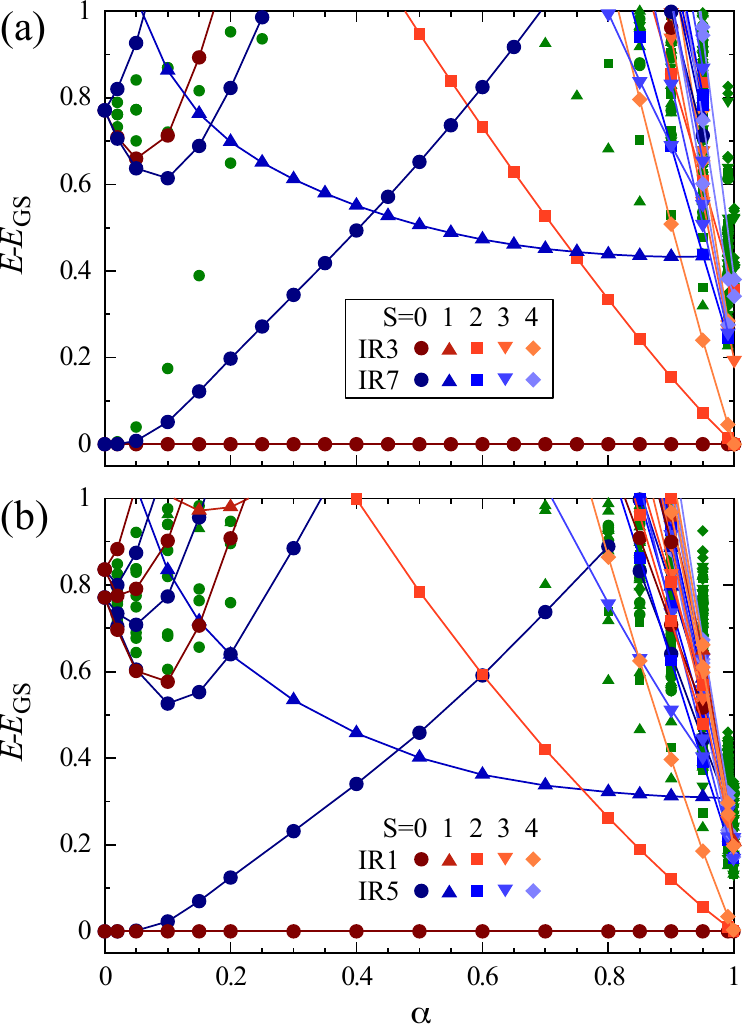}
\end{centering}
\caption{Energy spectra for two different clusters $N=27$ and 36. 
The level crossing occurs at $\alpha_{c}\approx0.6$ 
between the energy levels of the same quantum number for both the clusters, 
which indicates the possible quantum phase transition. 
All the energies are measured from the lowest energy singlet state. 
\label{fig:energy_spectra}}
\end{figure}
We now consider the ground state and lowest excited states over the full range $\alpha\in[0,1]$. 
Figure \ref{fig:energy_spectra} shows the energy spectra as a function of $\alpha$ for the $N=27$ and $36$ site clusters. 
At  $\alpha=0$, the excitation gap above the three fold degenerate ground state corresponds exactly to the value of $\Delta$ obtained by diagonalizing the Heisenberg chains in Sec III. 
Due to finite size effects, the three-fold degeneracy of the ground state is lifted immediately for any $\alpha>0$.
A level crossing in the lowest excitations is prominent near $\alpha=0.6$ for both clusters, indicating that the system undergoes a first order quantum phase transititon. \cite{sachdev2007,Zamphir02,Arias03}

\begin{figure}
 \begin{centering}
 \scalebox{1.0}{\includegraphics[width=80mm]{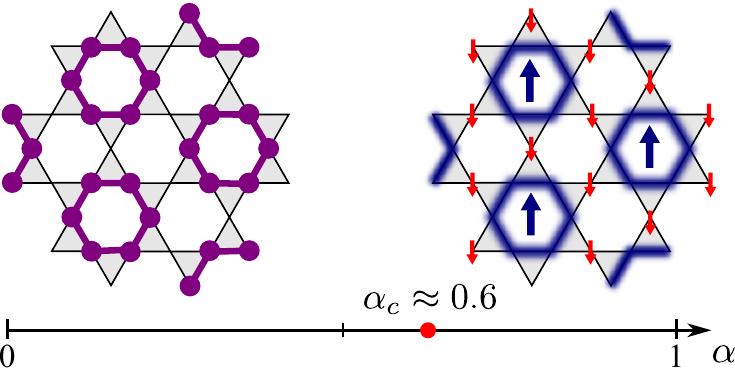}}
 \end{centering}
\caption{Ground state phase diagram of Eq.(\ref{eq:Hlow}). 
With increasing $\alpha$, the short-loop phase undergoes the first order quantum phase transition 
into the plaquette phase at around $\alpha \sim 0.6$. 
The solid hexagons on the upper left panel denote the short loops formed by the neighboring electrons, 
and the blurred hexagons on the upper right panel indicate the presence of the resonating plaquettes.}

\label{gspt}
\end{figure}

Combining the findings of the previous sections and the exact diagonalization data in Fig~\ref{fig:energy_spectra}, 
we reach the phase diagram shown in Fig.~\ref{gspt}. 
Two phases are separated by a first order phase transition; 
For $\alpha\alt 0.6$ we find the ``short loop'' phase, representing a charge ordered phase that has both a charge and a spin gap. 

\subsection{Correlation functions and structure factors}
We calculate several different kinds of correlation functions for the $N=36$ cluster which serve as characteristic finger prints of the phases.
\begin{figure}
 \begin{center}
\includegraphics[width=8.5truecm]{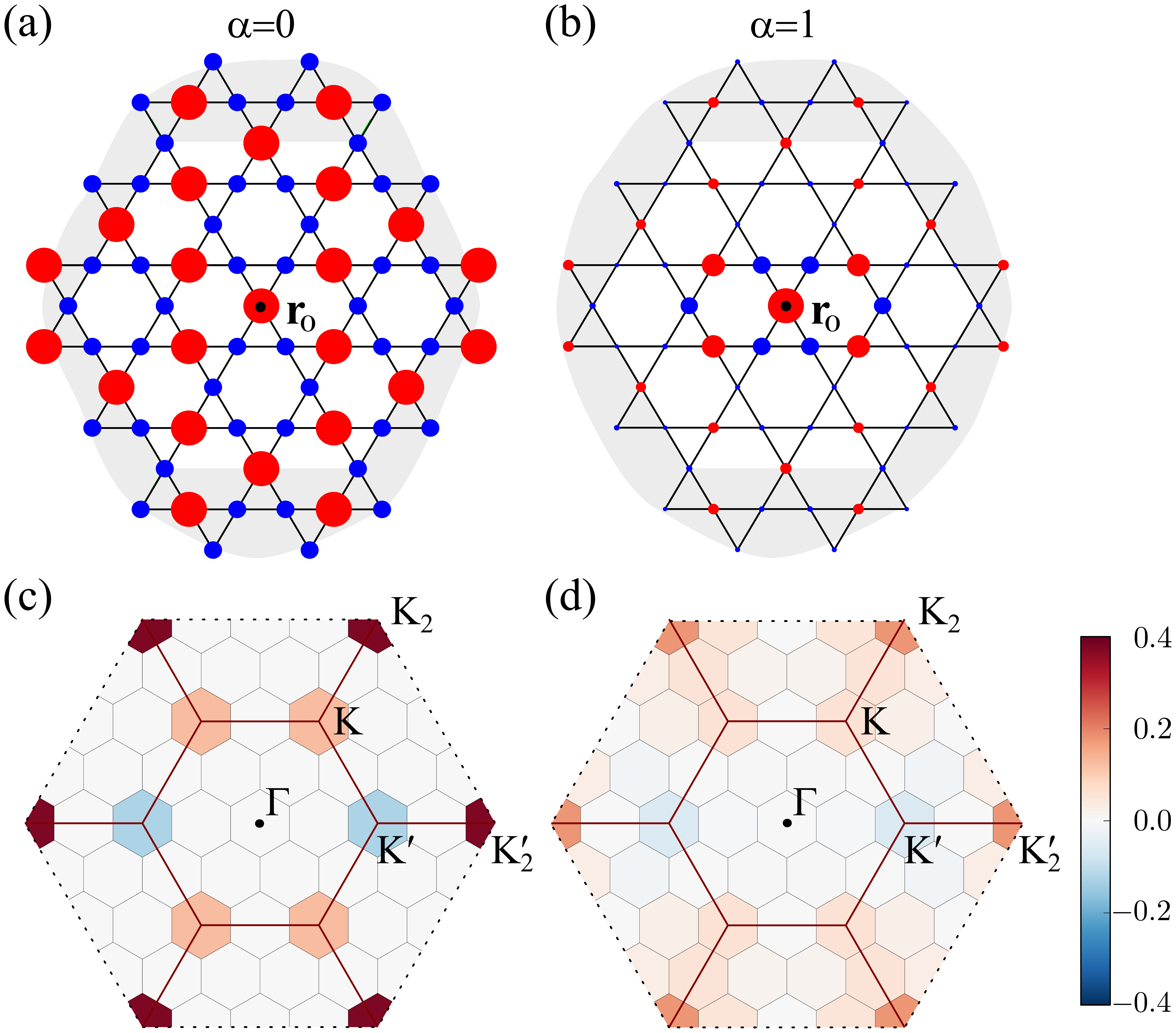}
 \end{center}
\caption{Panels (a) and (b) show the charge-charge correlation function in real space for $\alpha=0$ and 1, respectively. The correlations are calculated with respect to the marked site $\mathbf{r_0}$ at the center.
The radius of the dots is proportional to the absolute value of the correlation, 
while the colour encodes the sign 
(red/blue corresponds to positive/negative value). 
Panels (c) and (d) display the same data in momentum space in the extended Brillouin zone. 
The maxima in panel (c) is found at ${\bf Q} = (8\pi/3,0)$ and symmetry related points (high symmetry points denoted by $K_2$ and $K_2'$).
%, ($4\pi/3,4\pi/\surd3$), ($-4\pi/3,4\pi/\surd3$), ($-8\pi/3,0$), ($-4\pi/3,-4\pi/\surd3$) and ($4\pi/3,-4\pi/\surd3$). 
\label{fig:corr_density}
}
\end{figure} 
We begin by studying the charge-charge correlation functions, 
\begin{equation}
C^c(\mathbf{r}_0,\mathbf{r}_j)=\langle n({\mathbf{r}_0})n({\mathbf{r}_j})\rangle-\langle  n({\mathbf{r}_0)}\rangle\langle n({\mathbf{r}_j})\rangle,
\end{equation}
where $\mathbf{r}_{0}$ and $\mathbf{r}_{j}$ are the positions of charges on the kagome lattice and the expectation values are taken with respect to the ground state. 
The $n({\mathbf{r}_j})$ is the occupation number operator which measures whether the charge is present at $\mathbf{r}_j$ regardless of its spin orientation.
Figure~\ref{fig:corr_density}(a) shows the density plot of $C^c(\mathbf{r}_0,\mathbf{r}_j)$ at $\alpha=0$, 
which describes the charge order of the ``short loop phase''. 
The charges on the hexagons, which form short loops, are perfectly correlated. 
With increasing $\alpha$, this order gradually melts toward $\alpha=1$ at which only the short range correlations remain as shown in Fig.~\ref{fig:corr_density}(b).
The corresponding structure factor, 
\begin{equation}
S^c(\mathbf{q}) = \frac{1}{N}\sum_j  e^{-i\mathbf{q}\cdot(\mathbf{r}_j-\mathbf{r}_0)}\  C^c(\mathbf{r}_0,\mathbf{r}_j)\label{eq:s}
\end{equation}
is calculated in the extended Brillouin zone as shown in Fig.~\ref{fig:corr_density}. 
Note that we do not average the structure factor over the entire unit cell but instead calculate it for a specific center. 
Thus we have both positive an negative contributions.
In experiments, one would observe the averaged structure factor.
The ordering wave vectors $\bf{Q}$ of the ``short loop phase'' lie at the corners ($K_2$ and $K_2'$) of the extended Brillouin zone where the sharp peaks are observed. 
\begin{figure}
 \begin{center}
\includegraphics[width=8.5truecm]{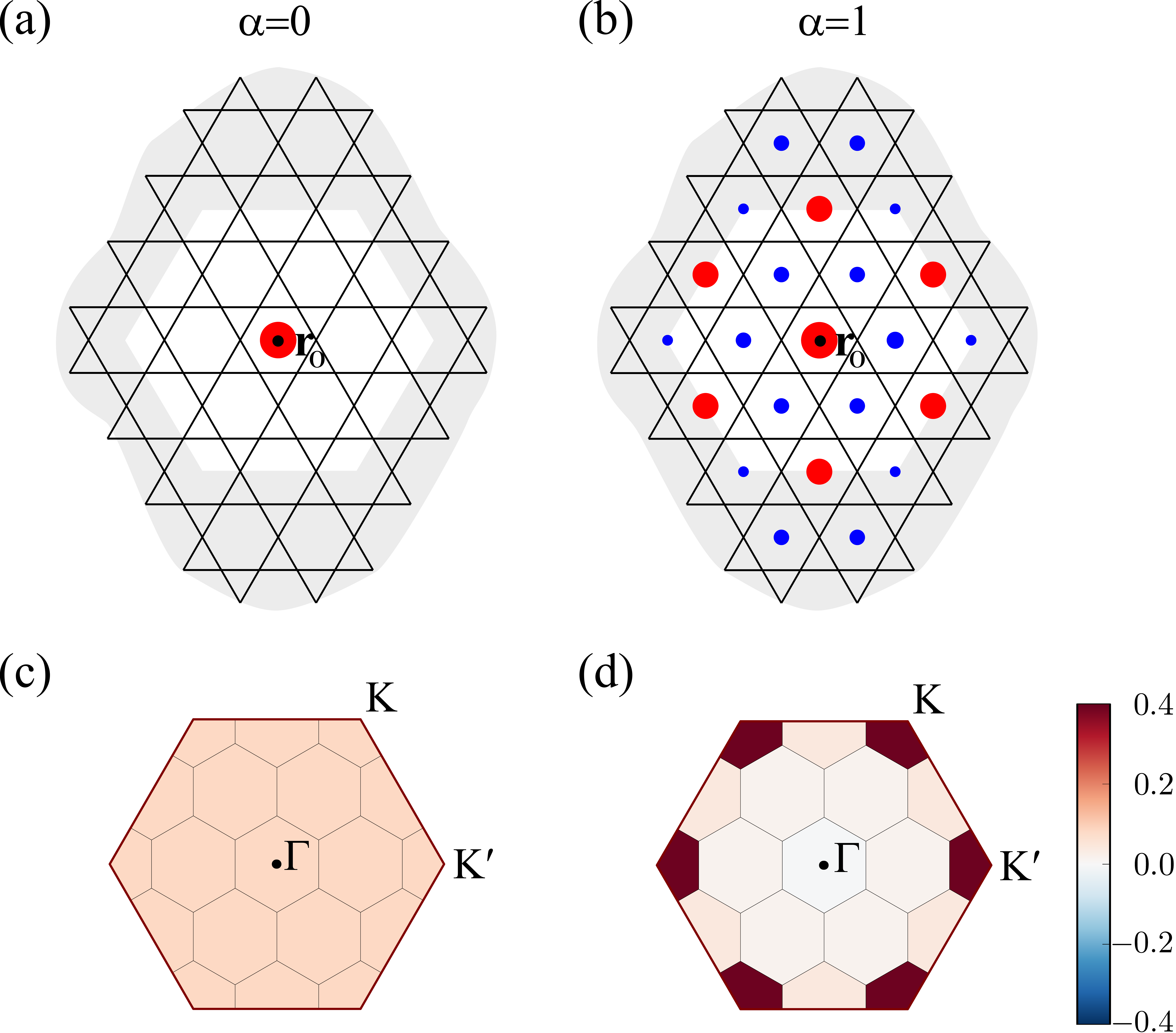}
 \end{center}
\caption{Panels (a) and (b) show the plaquette-plaquette correlation function in real space for $\alpha=0$ and 1, respectively. The correlations are calculated with respect to the marked site $\mathbf{r_0}$ at the center.
The radius of the dots is proportional to the absolute value of the correlation while the color encodes the sign 
(red/blue corresponds to positive/negative value). 
Panels (c) and (d) show the same data in momentum space in the first Brillouin zone. 
The peaks in panel (d) are found at $K$ and $K'$ points, with ${\bf Q} = (2\pi/3,2\pi/\sqrt{3})$ and ${\bf Q} = (4\pi/3,0)$, respectively. 
\label{fig:corr_plaq}
}
\end{figure} 

Next we consider the plaquette-plaquette correlation function 
\begin{equation}
C^h(\mathbf{R}_0,\mathbf{R}_j)=\langle h_{\hexagon}({\mathbf{R}_0})h_{\hexagon}({\mathbf{R}_j})\rangle-\langle h_{\hexagon}({\mathbf{R}_0)}\rangle\langle h_{\hexagon}({\mathbf{R}_j})\rangle,
\end{equation}
where the operators $h_{\hexagon}(\mathbf{R}_{0})$ and $h_{\hexagon}(\mathbf{R}_{j})$ are those representing the resonance, 
as defined in Eq.~(\ref{eq:ring}) on the hexagons centered at positions $\mathbf{R}_{0}$ and $\mathbf{R}_{j}$, respectively. 
The centers of these hexagons form a triangular lattice. 
At $\alpha=0$, $C^h(\mathbf{R}_0,\mathbf{R}_j)$ vanishes except for $\mathbf{R}_0 = \mathbf{R}_j$, as shown in Fig.~\ref{fig:corr_plaq}(a). 
This is because the charges are perfectly localized on short loops, and thus cannot resonate. 
By contrast, we find a clear sign of the plaquette ordering at $\alpha=1$ in Fig.~\ref{fig:corr_plaq}(b), 
whose spacial pattern are exactly the one expected in Fig.~\ref{gspt}. 
The structure factor $S^h(\bf{Q})$ for both cases is calculated analogously to Eq.~(\ref{eq:s}) 
and is displayed over the first Brillouin zone in Figs.~\ref{fig:corr_plaq}(c) and \ref{fig:corr_plaq}(d). 
The one at $\alpha=0$ is structureless, whereas at $\alpha=1$ we observe sharp peaks at the corners of the first Brillouin zone. 
\begin{figure}
 \begin{center}
\includegraphics[width=8.5truecm]{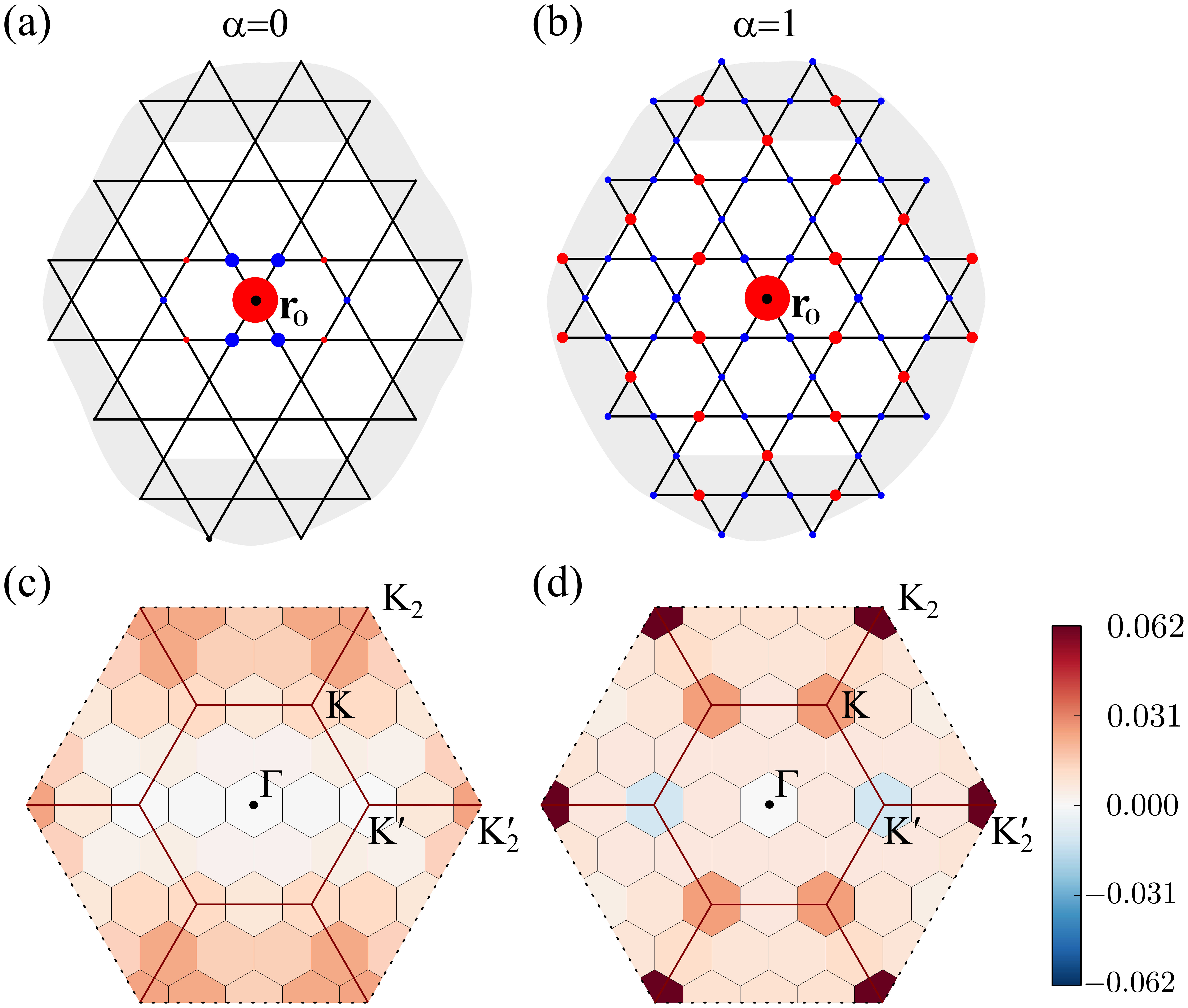}
 \end{center}
\caption{Panels (a) and (b) show the spin-spin correlation function in real space for $\alpha=0$ and 1, respectively. The correlations are calculated with respect to the marked site $\mathbf{r_0}$ at the center.
The radius of the dots is proportional to the absolute value of the correlation while the color encodes the sign 
(red/blue corresponds to positive/negative value). 
Panels (c) and (d) show the same data in momentum space in the extended Brillouin zone. 
\label{fig:corr_spin}
}
\end{figure} 
\begin{figure}
 \begin{centering}
\includegraphics[width=8truecm]{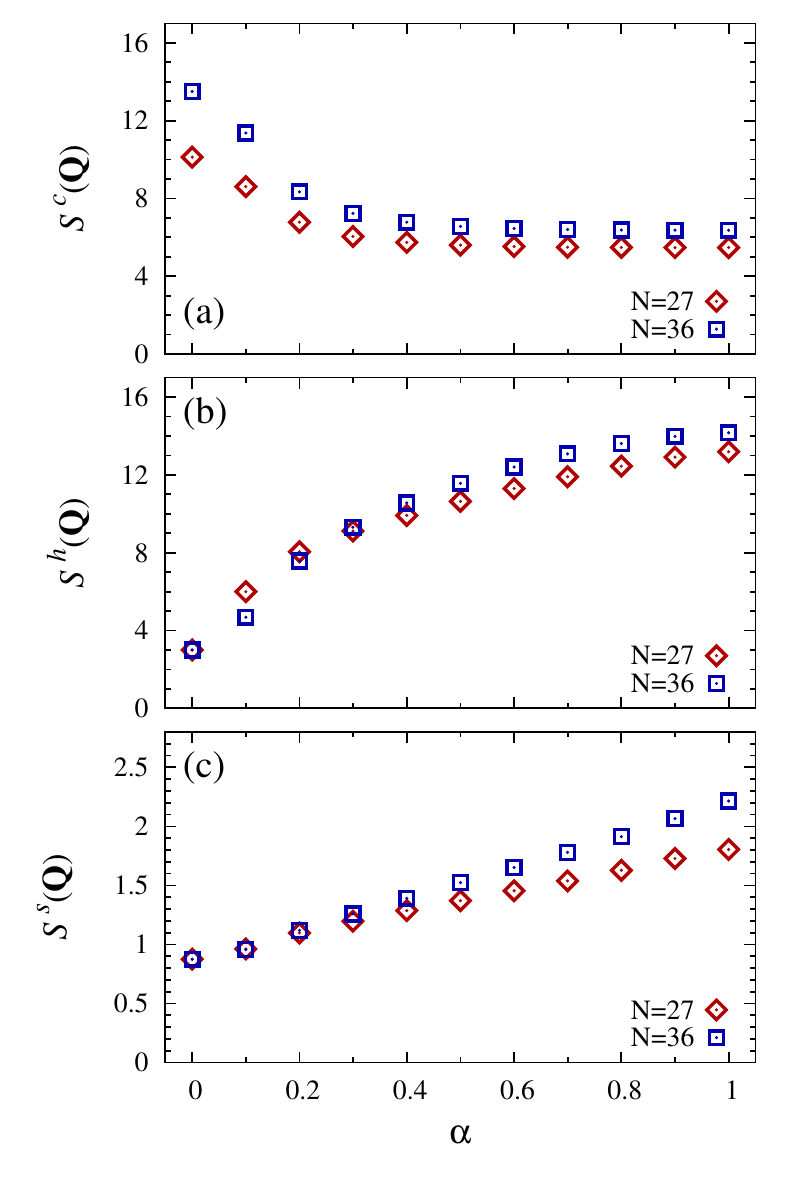}
 \end{centering}
\caption{Amplitudes of the (a) charge, (b) plaquette, (c) and spin structure factors, shown in Figs.\ref{fig:corr_density}-\ref{fig:corr_spin},
at the corresponding ordering wave vectors ${\bf Q}$. 
The gradual changes in all three panels indicate the melting of different kinds of ordering as the parameter $\alpha$ is varied. 
}
\label{melt-order}
\end{figure}

Finally we consider the spin-spin correlation function, 
\begin{equation}
C^s(\mathbf{r}_0,\mathbf{r}_j)=\langle \mathbf{S}({\mathbf{r}_0})\mathbf{S}({\mathbf{r}_j})\rangle-\langle \mathbf{S}({\mathbf{r}_0)}\rangle\langle \mathbf{S}({\mathbf{r}_j})\rangle, 
\end{equation}
where the operator $\mathbf{S}_{\mathbf{r}_j}$ is the spin-1/2 spin operator at site $\mathbf{r}_{j}$. 
For $\alpha=0$ the spins living on different short-loops are uncorrelated, and this is clearly seen in $C^s(\mathbf{r}_0,\mathbf{r}_j)$ which vanishes once the distances are $|\mathbf{r}_0-\mathbf{r}_j|>2$ [Fig.~\ref{fig:corr_spin}(a) and its corresponding structure factor in Fig.~\ref{fig:corr_spin}(c)].
The spin structure becomes more distinct as $\alpha$ goes to 1 [see Figs.\ref{fig:corr_spin}(b) and \ref{fig:corr_spin}(d)], 
and its textures in real and reciprocal spaces follow those of the charge in Figs.~\ref{fig:corr_density}(b) and \ref{fig:corr_density}(d). 
As we discussed earlier, the spins living on two dynamic sublattices form large ferromagnetic spins, $S_A$ and $S_B$, 
and one will find large correlation between the spins belonging to the same sublattice, if the dynamic sublattice could be extracted. 
However, in real space, the strong change fluctuations will cause the mixing of the two dynamic sublattices, 
and quite large part of the real space correlations are cancelled out. 
\par
 In Fig.~\ref{melt-order} we show how the amplitudes of different structure factors 
at the respective ordering wave vectors (the ${\bf Q}$-point which has largest amplitude of the structural factors) evolve. 
Here we multiply the amplitudes by the system size $N$, in order to compare the results of different size on the same ground, 
assuming that the sum rules are fulfilled. 
While the general tendency is clear, it is difficult to identify the phase transition point between the two phases in the correlation functions, presumably due to finite size effect.

%*%*%*%*%*%*%*%*%*%*%*%*  
%*%*%*%*%*%*%*%*%*%*%*%*  

\section{Summary and Outlook}\label{sum}

We considered  a system of strongly correlated electrons on kagome lattice at one--third filling, and focussed on the interplay of charge and spin fluctuations. 
A surprising aspect of our findings is that the originally complicated correlation and dynamics of the charge and spin degrees of freedom could be well separated within our approach in the strong coupling limit. We derived an effective  Hamiltonian which is acting on the low-energy manifold consisting of configurations with exactly two electrons per each triangle of the kagome lattice with no double occupancy. 

We discussed in detail two limiting cases: (i) In the limit where charge fluctuations dominate, a robust resonating ``plaquette ordered'' phase is found. The charge fluctuations conserve the magnetization on two dynamical sublattices and maximize the total spin in each sector,\cite{pollmann2008} yielding a huge degeneracy. Small spin fluctuations then couple two  giant spins weakly, leading to gapless spin excitations. (ii) In the limit where spin fluctuations dominate, the electrons are confined to short loops around the hexagons to maximize the energy gain due to spin fluctuations. 

Using large scale exact diagonalization, we evaluated the phase diagram 
and found a first order transition separating the ``plaquette ordered'' 
and the ``short loop'' phase. For both phases we obtained the fingerprints in form experimentally 
accessible signatures like spin and charge structure factors. 

To find the physics we discuss in this paper, we need to search for strongly correlated materials with a kagome lattice structure in the mixed valence regime. This might possibly be achieved by heavy doping  of the current kagome spin-liquid candidates like ZnCu$_3$(OH)$_6$Cl$_2$\cite{Mendels2010,Han2012} and Rb$_2$Cu$_3$SnF$_{12}$\cite{Morita2008} to get the desired filling factors.

\emph{Note added.} Upon completion of the present manuscript we learned about a similar work by K. Ferhat and A. Ralko, Ref.~\onlinecite{Ralko2013}: they considered the Hubbard model, Eq.~(\ref{eq:extended_hub}), without mapping to an effective model, and reached conclusions similar to ours for the considered limit.

\acknowledgments
The authors thank Peter Fulde and Andreas L\"auchli for stimulating discussions. This work is supported by Grant-in-Aid for Scientific Research (No. 25800204) from the Ministry of Education, Science, Sports and Culture of Japan and by the Hungarian OTKA Grant No. 106047. K.P. and C. H. gratefully acknowledge the hospitality of the guest program of MPI-PKS Dresden, where much of this work was carried out.

\appendix

\section{Proof for dynamic sublattice rule}
\label{sec:dynsublatt}
In this appendix we prove that the effective ring-exchange Hamiltonian ($\alpha=1$) conserves the magnetization on the dynamic sublattices. 
The rules for the ``dynamic sublattice'' are simple: (i) neighboring electrons have different sublattice labels and (ii) next-nearest neighbor electrons have the same sublattice label  on each hexagon of the kagome lattice. 
Once we have the configuration which fulfills (i) and (ii), the effective Hamiltonian ${\mathcal H}_{\rm ring}$ preserves this rule.
A flippable hexagon involves necessarily only loop segments of the same kind which is demanded by (ii). 
Also the protruding bonds coming out of a flippable hexagon are always exactly of the other kind to which the hexagon is made of. 
Let us assume without loss of generality an $A$-$A$-$A$ configuration on the flipped hexagon. 
Then the six protruding bonds will essentially be labelled by $B$. 
Flipping the loop segments around a hexagon does not change the positions relative to each other and thus (i) and (ii) are fulfilled in the resulting hexagon configuration. 
Next we have to check the six neighboring hexagons. 
The three hexagons which loose a loop segment are trivially fulfilling the conditions. 
For the hexagons which gain a loop segment, we need to argue a little more. 
The neighboring links of the added segment are necessarily occupied by a $B$ segment since the initial configuration was a valid closely-packed loop configuration and it fulfilled (i). 
Since the neighboring segments are of the $B$ type, we know that the next-nerest neighboring segments are of the $A$ type or empty. Thus the resulting configurations fulfills (i) 
and (ii).

\section{Irreducible representations of the 36 site cluster}
\label{sec:IR}

\begin{figure}
 \begin{centering}
\includegraphics[width=6.truecm]{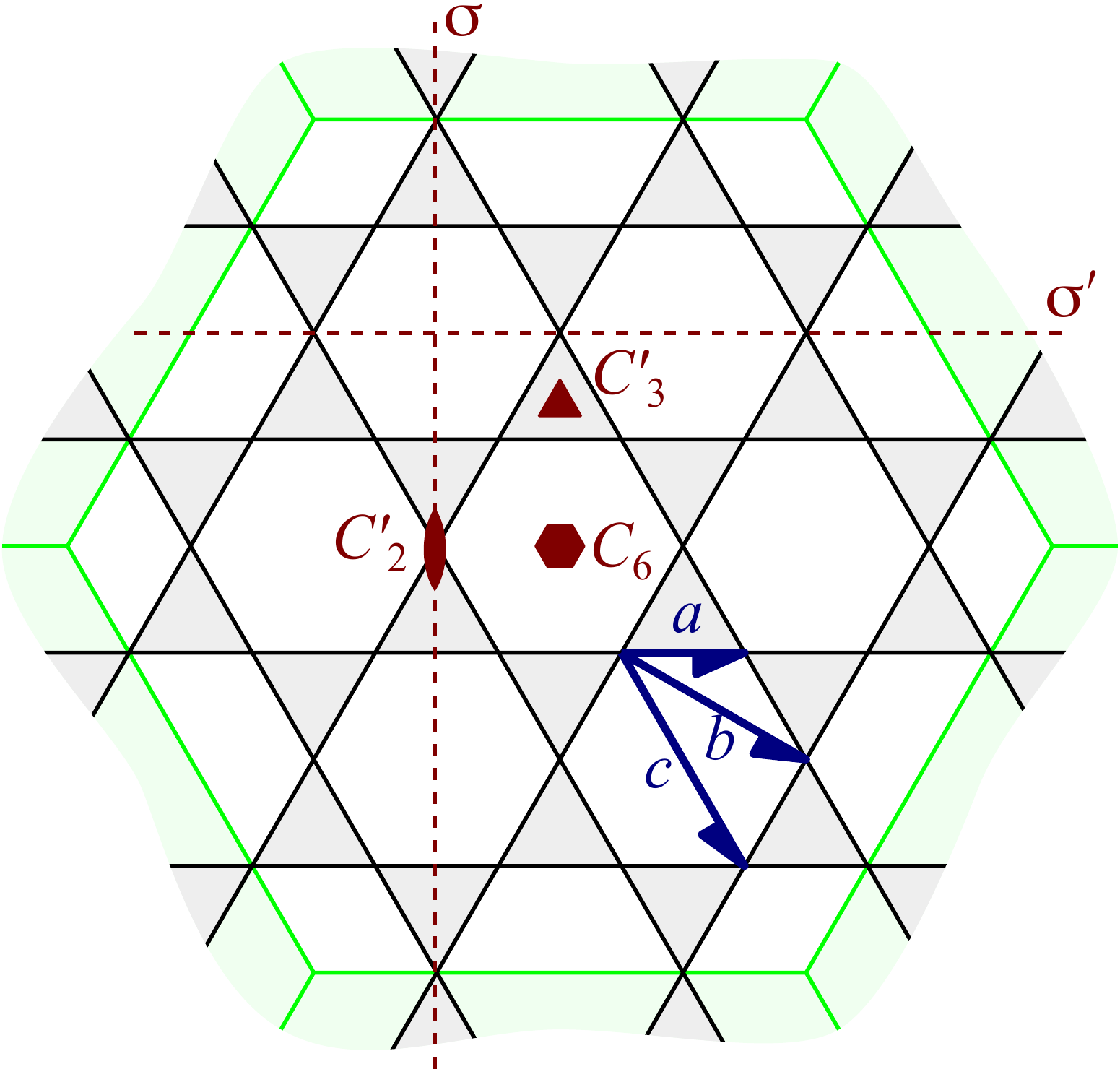}
 \end{centering}
\caption{Description of the $N=36$ site cluster with periodic boundary conditions. The symmetries of the point group are denoted by dark red: the hexagon denotes the rotation center of order six ($C_6$), the triangle the rotation center of order three ($C_3'$), while the oval the rotation center of order 2 ($C_2'$). Dashed lines are the reflexions $\sigma$ and $\sigma'$. Dark blue lines with half arrow denote glide reflections.
\label{fig:36cluster}}
\end{figure}

The symmetry group elements of the 36 site cluster with periodic boundary conditions are shown in Fig.~\ref{fig:36cluster}. The symmetry group of the kagome lattice is the wallpaper group {\bf p6m}, with the point group ${\mathcal D_6}$. The 36 site cluster consists of 12 unit cells, and since the order of the ${\mathcal D_6}$ is 12, the total number of the symmetry group elements is $12\times12=144$.

\begin{widetext}
The character table for the 36 site cluster is as follows:
\begin{equation}
\begin{array}{r|rrrrrrrrrrrrrrr|r}
%\text{IR} & I & 6g_{2} & 18g_{3} & 9g_{4} & 3g_{5} & 6g_{6} & 12g_{7} & 12g_{8} & 18g_{9} & 6g_{10} & 3g_{11} & 2g_{12} & 24g_{13} & 16g_{14} & 8g_{15}  & \text{BZ}\\
\text{IR} & I & 6\sigma & 18\sigma' & 9C_2' & 3C_6^3 & 6T_1 & 12c & 12a & 18b & 6a^3 & 3T_1 T_2 & 2T_1^2 & 24C_6 & 16C_3' & 8C_6^2  & \text{BZ}\\
\hline
 1 & 1 & 1 & 1 & 1 & 1 & 1 & 1 & 1 & 1 & 1 & 1 & 1 & 1 & 1 & 1  & \Gamma\\
 2 & 1 & 1 & -1 & -1 & -1 & 1 & 1 & 1 & -1 & 1 & 1 & 1 & -1 & 1 & 1 & \Gamma\\
 3 & 1 & -1 & 1 & -1 & -1 & 1 & -1 & -1 & 1 & -1 & 1 & 1 & -1 & 1 & 1 & \Gamma\\
 4 & 1 & -1 & -1 & 1 & 1 & 1 & -1 & -1 & -1 & -1 & 1 & 1 & 1 & 1 & 1 & \Gamma\\
\hline
 5 & 2 & 2 & 0 & 0 & 0 & -1 & -1 & -1 & 0 & 2 & 2 & -1 & 0 & -1 & 2 & K\\
 6 & 2 & -2 & 0 & 0 & 0 & -1 & 1 & 1 & 0 & -2 & 2 & -1 & 0 & -1 & 2 & K \\
\hline
 7 & 2 & 0 & 0 & 2 & 2 & 2 & 0 & 0 & 0 & 0 & 2 & 2 & -1 & -1 & -1 & \Gamma\\
 8 & 2 & 0 & 0 & -2 & -2 & 2 & 0 & 0 & 0 & 0 & 2 & 2 & 1 & -1 & -1 & \Gamma\\
\hline
 9 & 4 & 0 & 0 & 0 & 0 & -2 & 0 & 0 & 0 & 0 & 4 & -2 & 0 & 1 & -2 & K\\
\hline
10 & 3 & 1 & 1 & -1 & 3 & -1 & 1 & -1 & -1 & -1 & -1 & 3 & 0 & 0 & 0 & M\\
11 & 3 & 1 & -1 & 1 & -3 & -1 & 1 & -1 & 1 & -1 & -1 & 3 & 0 & 0 & 0 & M\\
12 & 3 & -1 & 1 & 1 & -3 & -1 & -1 & 1 & -1 & 1 & -1 & 3 & 0 & 0 & 0 & M\\
13 & 3 & -1 & -1 & -1 & 3 & -1 & -1 & 1 & 1 & 1 & -1 & 3 & 0 & 0 & 0 & M\\
\hline
14 & 6 & 2 & 0 & 0 & 0 & 1 & -1 & 1 & 0 & -2 & -2 & -3 & 0 & 0 & 0 \\
15 & 6 & -2 & 0 & 0 & 0 & 1 & 1 & -1 & 0 & 2 & -2 & -3 & 0 & 0 & 0 \\
\end{array}
\end{equation}
where we also specified the point(s) in the Brillouin zone it belongs to.
\end{widetext}

%\bibliographystyle{apsrev4-1}
%\bibliography{paper_KRC_library}

\begin{thebibliography}{36}%
\makeatletter
\providecommand \@ifxundefined [1]{%
 \@ifx{#1\undefined}
}%
\providecommand \@ifnum [1]{%
 \ifnum #1\expandafter \@firstoftwo
 \else \expandafter \@secondoftwo
 \fi
}%
\providecommand \@ifx [1]{%
 \ifx #1\expandafter \@firstoftwo
 \else \expandafter \@secondoftwo
 \fi
}%
\providecommand \natexlab [1]{#1}%
\providecommand \enquote  [1]{``#1''}%
\providecommand \bibnamefont  [1]{#1}%
\providecommand \bibfnamefont [1]{#1}%
\providecommand \citenamefont [1]{#1}%
\providecommand \href@noop [0]{\@secondoftwo}%
\providecommand \href [0]{\begingroup \@sanitize@url \@href}%
\providecommand \@href[1]{\@@startlink{#1}\@@href}%
\providecommand \@@href[1]{\endgroup#1\@@endlink}%
\providecommand \@sanitize@url [0]{\catcode `\\12\catcode `\$12\catcode
  `\&12\catcode `\#12\catcode `\^12\catcode `\_12\catcode `\%12\relax}%
\providecommand \@@startlink[1]{}%
\providecommand \@@endlink[0]{}%
\providecommand \url  [0]{\begingroup\@sanitize@url \@url }%
\providecommand \@url [1]{\endgroup\@href {#1}{\urlprefix }}%
\providecommand \urlprefix  [0]{URL }%
\providecommand \Eprint [0]{\href }%
\providecommand \doibase [0]{http://dx.doi.org/}%
\providecommand \selectlanguage [0]{\@gobble}%
\providecommand \bibinfo  [0]{\@secondoftwo}%
\providecommand \bibfield  [0]{\@secondoftwo}%
\providecommand \translation [1]{[#1]}%
\providecommand \BibitemOpen [0]{}%
\providecommand \bibitemStop [0]{}%
\providecommand \bibitemNoStop [0]{.\EOS\space}%
\providecommand \EOS [0]{\spacefactor3000\relax}%
\providecommand \BibitemShut  [1]{\csname bibitem#1\endcsname}%
\let\auto@bib@innerbib\@empty
%</preamble>
\bibitem [{\citenamefont {Castelnovo}\ \emph {et~al.}(2008)\citenamefont
  {Castelnovo}, \citenamefont {Moessner},\ and\ \citenamefont
  {Sondhi}}]{moessner}%
  \BibitemOpen
  \bibfield  {author} {\bibinfo {author} {\bibfnamefont {C.}~\bibnamefont
  {Castelnovo}}, \bibinfo {author} {\bibfnamefont {R.}~\bibnamefont
  {Moessner}}, \ and\ \bibinfo {author} {\bibfnamefont {S.~L.}\ \bibnamefont
  {Sondhi}},\ }\href@noop {} {\bibfield  {journal} {\bibinfo  {journal}
  {Nature}\ }\textbf {\bibinfo {volume} {451}},\ \bibinfo {pages} {42}
  (\bibinfo {year} {2008})}\BibitemShut {NoStop}%
\bibitem [{\citenamefont {Moessner}\ and\ \citenamefont
  {Sondhi}(2003)}]{moessner2003}%
  \BibitemOpen
  \bibfield  {author} {\bibinfo {author} {\bibfnamefont {R.}~\bibnamefont
  {Moessner}}\ and\ \bibinfo {author} {\bibfnamefont {S.~L.}\ \bibnamefont
  {Sondhi}},\ }\href {\doibase 10.1103/PhysRevB.68.184512} {\bibfield
  {journal} {\bibinfo  {journal} {Phys. Rev. B}\ }\textbf {\bibinfo {volume}
  {68}},\ \bibinfo {pages} {184512} (\bibinfo {year} {2003})}\BibitemShut
  {NoStop}%
\bibitem [{\citenamefont {Hermele}\ \emph {et~al.}(2004)\citenamefont
  {Hermele}, \citenamefont {Fisher},\ and\ \citenamefont
  {Balents}}]{hermele2004}%
  \BibitemOpen
  \bibfield  {author} {\bibinfo {author} {\bibfnamefont {M.}~\bibnamefont
  {Hermele}}, \bibinfo {author} {\bibfnamefont {M.~P.~A.}\ \bibnamefont
  {Fisher}}, \ and\ \bibinfo {author} {\bibfnamefont {L.}~\bibnamefont
  {Balents}},\ }\href {\doibase 10.1103/PhysRevB.69.064404} {\bibfield
  {journal} {\bibinfo  {journal} {Phys. Rev. B}\ }\textbf {\bibinfo {volume}
  {69}},\ \bibinfo {pages} {064404} (\bibinfo {year} {2004})}\BibitemShut
  {NoStop}%
\bibitem [{\citenamefont {Shannon}\ \emph {et~al.}(2012)\citenamefont
  {Shannon}, \citenamefont {Sikora}, \citenamefont {Pollmann}, \citenamefont
  {Penc},\ and\ \citenamefont {Fulde}}]{sikora2012}%
  \BibitemOpen
  \bibfield  {author} {\bibinfo {author} {\bibfnamefont {N.}~\bibnamefont
  {Shannon}}, \bibinfo {author} {\bibfnamefont {O.}~\bibnamefont {Sikora}},
  \bibinfo {author} {\bibfnamefont {F.}~\bibnamefont {Pollmann}}, \bibinfo
  {author} {\bibfnamefont {K.}~\bibnamefont {Penc}}, \ and\ \bibinfo {author}
  {\bibfnamefont {P.}~\bibnamefont {Fulde}},\ }\href {\doibase
  10.1103/PhysRevLett.108.067204} {\bibfield  {journal} {\bibinfo  {journal}
  {Phys. Rev. Lett.}\ }\textbf {\bibinfo {volume} {108}},\ \bibinfo {pages}
  {067204} (\bibinfo {year} {2012})}\BibitemShut {NoStop}%
\bibitem [{\citenamefont {Benton}\ \emph {et~al.}(2012)\citenamefont {Benton},
  \citenamefont {Sikora},\ and\ \citenamefont {Shannon}}]{Benton2012}%
  \BibitemOpen
  \bibfield  {author} {\bibinfo {author} {\bibfnamefont {O.}~\bibnamefont
  {Benton}}, \bibinfo {author} {\bibfnamefont {O.}~\bibnamefont {Sikora}}, \
  and\ \bibinfo {author} {\bibfnamefont {N.}~\bibnamefont {Shannon}},\ }\href
  {\doibase 10.1103/PhysRevB.86.075154} {\bibfield  {journal} {\bibinfo
  {journal} {Phys. Rev. B}\ }\textbf {\bibinfo {volume} {86}},\ \bibinfo
  {pages} {075154} (\bibinfo {year} {2012})}\BibitemShut {NoStop}%
\bibitem [{\citenamefont {Mendels}\ and\ \citenamefont
  {Wills}(2011)}]{Mendels2011}%
  \BibitemOpen
  \bibfield  {author} {\bibinfo {author} {\bibfnamefont {P.}~\bibnamefont
  {Mendels}}\ and\ \bibinfo {author} {\bibfnamefont {A.~S.}\ \bibnamefont
  {Wills}},\ }in\ \href {\doibase 10.1007/978-3-642-10589-0_9} {\emph {\bibinfo
  {booktitle} {Introduction to Frustrated Magnetism}}},\ \bibinfo {series}
  {Springer Series in Solid-State Sciences}, Vol.\ \bibinfo {volume} {164},\
  \bibinfo {editor} {edited by\ \bibinfo {editor} {\bibfnamefont
  {C.}~\bibnamefont {Lacroix}}, \bibinfo {editor} {\bibfnamefont
  {P.}~\bibnamefont {Mendels}}, \ and\ \bibinfo {editor} {\bibfnamefont
  {F.}~\bibnamefont {Mila}}}\ (\bibinfo  {publisher} {Springer Berlin
  Heidelberg},\ \bibinfo {year} {2011})\ pp.\ \bibinfo {pages}
  {207--238}\BibitemShut {NoStop}%
\bibitem [{\citenamefont {Yan}\ \emph {et~al.}(2011)\citenamefont {Yan},
  \citenamefont {Huse},\ and\ \citenamefont {White}}]{white}%
  \BibitemOpen
  \bibfield  {author} {\bibinfo {author} {\bibfnamefont {S.}~\bibnamefont
  {Yan}}, \bibinfo {author} {\bibfnamefont {D.~A.}\ \bibnamefont {Huse}}, \
  and\ \bibinfo {author} {\bibfnamefont {S.~R.}\ \bibnamefont {White}},\
  }\href@noop {} {\bibfield  {journal} {\bibinfo  {journal} {Science}\ }\textbf
  {\bibinfo {volume} {332}},\ \bibinfo {pages} {1173} (\bibinfo {year}
  {2011})}\BibitemShut {NoStop}%
\bibitem [{\citenamefont {Depenbrock}\ \emph {et~al.}(2012)\citenamefont
  {Depenbrock}, \citenamefont {McCulloch},\ and\ \citenamefont
  {Schollwoeck}}]{depenbrock2012}%
  \BibitemOpen
  \bibfield  {author} {\bibinfo {author} {\bibfnamefont {S.}~\bibnamefont
  {Depenbrock}}, \bibinfo {author} {\bibfnamefont {I.~P.}\ \bibnamefont
  {McCulloch}}, \ and\ \bibinfo {author} {\bibfnamefont {U.}~\bibnamefont
  {Schollwoeck}},\ }\href@noop {} {\bibfield  {journal} {\bibinfo  {journal}
  {Physical Review Letters}\ }\textbf {\bibinfo {volume} {109}},\ \bibinfo
  {pages} {067201} (\bibinfo {year} {2012})}\BibitemShut {NoStop}%
\bibitem [{\citenamefont {Jiang}\ \emph {et~al.}(2012)\citenamefont {Jiang},
  \citenamefont {Wang},\ and\ \citenamefont {Balents}}]{balents2012}%
  \BibitemOpen
  \bibfield  {author} {\bibinfo {author} {\bibfnamefont {H.-C.}\ \bibnamefont
  {Jiang}}, \bibinfo {author} {\bibfnamefont {Z.}~\bibnamefont {Wang}}, \ and\
  \bibinfo {author} {\bibfnamefont {L.}~\bibnamefont {Balents}},\ }\href@noop
  {} {\bibfield  {journal} {\bibinfo  {journal} {Nature Physics}\ }\textbf
  {\bibinfo {volume} {8}},\ \bibinfo {pages} {902} (\bibinfo {year}
  {2012})}\BibitemShut {NoStop}%
\bibitem [{\citenamefont {Nishimoto}\ \emph {et~al.}(2013)\citenamefont
  {Nishimoto}, \citenamefont {Shibata},\ and\ \citenamefont
  {Hotta}}]{Nishimoto2013}%
  \BibitemOpen
  \bibfield  {author} {\bibinfo {author} {\bibfnamefont {S.}~\bibnamefont
  {Nishimoto}}, \bibinfo {author} {\bibfnamefont {N.}~\bibnamefont {Shibata}},
  \ and\ \bibinfo {author} {\bibfnamefont {C.}~\bibnamefont {Hotta}},\
  }\href@noop {} {\bibfield  {journal} {\bibinfo  {journal} {Nat. Commun.}\
  }\textbf {\bibinfo {volume} {4}} (\bibinfo {year} {2013})}\BibitemShut
  {NoStop}%
\bibitem [{\citenamefont {Fulde}\ \emph {et~al.}(2002)\citenamefont {Fulde},
  \citenamefont {Penc},\ and\ \citenamefont {Shannon}}]{fulde2001}%
  \BibitemOpen
  \bibfield  {author} {\bibinfo {author} {\bibfnamefont {P.}~\bibnamefont
  {Fulde}}, \bibinfo {author} {\bibfnamefont {K.}~\bibnamefont {Penc}}, \ and\
  \bibinfo {author} {\bibfnamefont {N.}~\bibnamefont {Shannon}},\ }\href@noop
  {} {\bibfield  {journal} {\bibinfo  {journal} {Annalen der Physik}\ }\textbf
  {\bibinfo {volume} {11}},\ \bibinfo {pages} {892} (\bibinfo {year}
  {2002})}\BibitemShut {NoStop}%
\bibitem [{\citenamefont {Runge}\ and\ \citenamefont
  {Fulde}(2004)}]{runge2004}%
  \BibitemOpen
  \bibfield  {author} {\bibinfo {author} {\bibfnamefont {E.}~\bibnamefont
  {Runge}}\ and\ \bibinfo {author} {\bibfnamefont {P.}~\bibnamefont {Fulde}},\
  }\href {\doibase 10.1103/PhysRevB.70.245113} {\bibfield  {journal} {\bibinfo
  {journal} {Phys. Rev. B}\ }\textbf {\bibinfo {volume} {70}},\ \bibinfo
  {pages} {245113} (\bibinfo {year} {2004})}\BibitemShut {NoStop}%
\bibitem [{\citenamefont {Pollmann}\ and\ \citenamefont
  {Fulde}(2006)}]{pollmann2006}%
  \BibitemOpen
  \bibfield  {author} {\bibinfo {author} {\bibfnamefont {F.}~\bibnamefont
  {Pollmann}}\ and\ \bibinfo {author} {\bibfnamefont {P.}~\bibnamefont
  {Fulde}},\ }\href@noop {} {\bibfield  {journal} {\bibinfo  {journal} {EPL
  (Europhysics Letters)}\ }\textbf {\bibinfo {volume} {75}},\ \bibinfo {pages}
  {133} (\bibinfo {year} {2006})}\BibitemShut {NoStop}%
\bibitem [{\citenamefont {O'Brien}\ \emph {et~al.}(2010)\citenamefont
  {O'Brien}, \citenamefont {Pollmann},\ and\ \citenamefont
  {Fulde}}]{obrien2010}%
  \BibitemOpen
  \bibfield  {author} {\bibinfo {author} {\bibfnamefont {A.}~\bibnamefont
  {O'Brien}}, \bibinfo {author} {\bibfnamefont {F.}~\bibnamefont {Pollmann}}, \
  and\ \bibinfo {author} {\bibfnamefont {P.}~\bibnamefont {Fulde}},\ }\href
  {\doibase 10.1103/PhysRevB.81.235115} {\bibfield  {journal} {\bibinfo
  {journal} {Phys. Rev. B}\ }\textbf {\bibinfo {volume} {81}},\ \bibinfo
  {pages} {235115} (\bibinfo {year} {2010})}\BibitemShut {NoStop}%
\bibitem [{\citenamefont {Poilblanc}\ \emph {et~al.}(2007)\citenamefont
  {Poilblanc}, \citenamefont {Penc},\ and\ \citenamefont
  {Shannon}}]{Poilblanc2007}%
  \BibitemOpen
  \bibfield  {author} {\bibinfo {author} {\bibfnamefont {D.}~\bibnamefont
  {Poilblanc}}, \bibinfo {author} {\bibfnamefont {K.}~\bibnamefont {Penc}}, \
  and\ \bibinfo {author} {\bibfnamefont {N.}~\bibnamefont {Shannon}},\ }\href
  {\doibase 10.1103/PhysRevB.75.220503} {\bibfield  {journal} {\bibinfo
  {journal} {Phys. Rev. B}\ }\textbf {\bibinfo {volume} {75}},\ \bibinfo
  {pages} {220503} (\bibinfo {year} {2007})}\BibitemShut {NoStop}%
\bibitem [{\citenamefont {Trousselet}\ \emph {et~al.}(2008)\citenamefont
  {Trousselet}, \citenamefont {Poilblanc},\ and\ \citenamefont
  {Moessner}}]{Trousselet2008}%
  \BibitemOpen
  \bibfield  {author} {\bibinfo {author} {\bibfnamefont {F.}~\bibnamefont
  {Trousselet}}, \bibinfo {author} {\bibfnamefont {D.}~\bibnamefont
  {Poilblanc}}, \ and\ \bibinfo {author} {\bibfnamefont {R.}~\bibnamefont
  {Moessner}},\ }\href {\doibase 10.1103/PhysRevB.78.195101} {\bibfield
  {journal} {\bibinfo  {journal} {Phys. Rev. B}\ }\textbf {\bibinfo {volume}
  {78}},\ \bibinfo {pages} {195101} (\bibinfo {year} {2008})}\BibitemShut
  {NoStop}%
\bibitem [{\citenamefont {Wen}\ \emph {et~al.}(2010)\citenamefont {Wen},
  \citenamefont {R{\"u}egg}, \citenamefont {Wang},\ and\ \citenamefont
  {Fiete}}]{wen2010}%
  \BibitemOpen
  \bibfield  {author} {\bibinfo {author} {\bibfnamefont {J.}~\bibnamefont
  {Wen}}, \bibinfo {author} {\bibfnamefont {A.}~\bibnamefont {R{\"u}egg}},
  \bibinfo {author} {\bibfnamefont {C.-C.~J.}\ \bibnamefont {Wang}}, \ and\
  \bibinfo {author} {\bibfnamefont {G.~A.}\ \bibnamefont {Fiete}},\ }\href
  {\doibase 10.1103/PhysRevB.82.075125} {\bibfield  {journal} {\bibinfo
  {journal} {Phys. Rev. B}\ }\textbf {\bibinfo {volume} {82}},\ \bibinfo
  {pages} {075125} (\bibinfo {year} {2010})}\BibitemShut {NoStop}%
\bibitem [{\citenamefont {Indergand}\ \emph {et~al.}(2006)\citenamefont
  {Indergand}, \citenamefont {L{\"a}uchli}, \citenamefont {Capponi},\ and\
  \citenamefont {Sigrist}}]{Indergand2006}%
  \BibitemOpen
  \bibfield  {author} {\bibinfo {author} {\bibfnamefont {M.}~\bibnamefont
  {Indergand}}, \bibinfo {author} {\bibfnamefont {A.}~\bibnamefont
  {L{\"a}uchli}}, \bibinfo {author} {\bibfnamefont {S.}~\bibnamefont
  {Capponi}}, \ and\ \bibinfo {author} {\bibfnamefont {M.}~\bibnamefont
  {Sigrist}},\ }\href {\doibase 10.1103/PhysRevB.74.064429} {\bibfield
  {journal} {\bibinfo  {journal} {Phys. Rev. B}\ }\textbf {\bibinfo {volume}
  {74}},\ \bibinfo {pages} {064429} (\bibinfo {year} {2006})}\BibitemShut
  {NoStop}%
\bibitem [{\citenamefont {Pollmann}\ \emph {et~al.}(2008)\citenamefont
  {Pollmann}, \citenamefont {Fulde},\ and\ \citenamefont
  {Shtengel}}]{pollmann2008}%
  \BibitemOpen
  \bibfield  {author} {\bibinfo {author} {\bibfnamefont {F.}~\bibnamefont
  {Pollmann}}, \bibinfo {author} {\bibfnamefont {P.}~\bibnamefont {Fulde}}, \
  and\ \bibinfo {author} {\bibfnamefont {K.}~\bibnamefont {Shtengel}},\
  }\href@noop {} {\bibfield  {journal} {\bibinfo  {journal} {Physical review
  letters}\ }\textbf {\bibinfo {volume} {100}},\ \bibinfo {pages} {136404}
  (\bibinfo {year} {2008})}\BibitemShut {NoStop}%
\bibitem [{\citenamefont {Anderson}(1956)}]{Anderson1956}%
  \BibitemOpen
  \bibfield  {author} {\bibinfo {author} {\bibfnamefont {P.~W.}\ \bibnamefont
  {Anderson}},\ }\href {\doibase 10.1103/PhysRev.102.1008} {\bibfield
  {journal} {\bibinfo  {journal} {Phys. Rev.}\ }\textbf {\bibinfo {volume}
  {102}},\ \bibinfo {pages} {1008} (\bibinfo {year} {1956})}\BibitemShut
  {NoStop}%
\bibitem [{\citenamefont {Fisher}\ and\ \citenamefont
  {Stephenson}(1963)}]{fisher}%
  \BibitemOpen
  \bibfield  {author} {\bibinfo {author} {\bibfnamefont {M.~E.}\ \bibnamefont
  {Fisher}}\ and\ \bibinfo {author} {\bibfnamefont {J.}~\bibnamefont
  {Stephenson}},\ }\href@noop {} {\bibfield  {journal} {\bibinfo  {journal}
  {Physical Review}\ }\textbf {\bibinfo {volume} {132}},\ \bibinfo {pages}
  {1411} (\bibinfo {year} {1963})}\BibitemShut {NoStop}%
\bibitem [{\citenamefont {Penc}\ \emph {et~al.}(1996)\citenamefont {Penc},
  \citenamefont {Shiba}, \citenamefont {Mila},\ and\ \citenamefont
  {Tsukagoshi}}]{penc96}%
  \BibitemOpen
  \bibfield  {author} {\bibinfo {author} {\bibfnamefont {K.}~\bibnamefont
  {Penc}}, \bibinfo {author} {\bibfnamefont {H.}~\bibnamefont {Shiba}},
  \bibinfo {author} {\bibfnamefont {F.}~\bibnamefont {Mila}}, \ and\ \bibinfo
  {author} {\bibfnamefont {T.}~\bibnamefont {Tsukagoshi}},\ }\href {\doibase
  10.1103/PhysRevB.54.4056} {\bibfield  {journal} {\bibinfo  {journal} {Phys.
  Rev. B}\ }\textbf {\bibinfo {volume} {54}},\ \bibinfo {pages} {4056}
  (\bibinfo {year} {1996})}\BibitemShut {NoStop}%
\bibitem [{\citenamefont {Mielke}(1992)}]{Mielke1992}%
  \BibitemOpen
  \bibfield  {author} {\bibinfo {author} {\bibfnamefont {A.}~\bibnamefont
  {Mielke}},\ }\href@noop {} {\bibfield  {journal} {\bibinfo  {journal}
  {Journal of Physics A: Mathematical and General}\ }\textbf {\bibinfo {volume}
  {25}},\ \bibinfo {pages} {4335} (\bibinfo {year} {1992})}\BibitemShut
  {NoStop}%
\bibitem [{\citenamefont {Mielke}\ and\ \citenamefont
  {Tasaki}(1993)}]{MielkeTasaki1993}%
  \BibitemOpen
  \bibfield  {author} {\bibinfo {author} {\bibfnamefont {A.}~\bibnamefont
  {Mielke}}\ and\ \bibinfo {author} {\bibfnamefont {H.}~\bibnamefont
  {Tasaki}},\ }\href {\doibase 10.1007/BF02108079} {\bibfield  {journal}
  {\bibinfo  {journal} {Communications in Mathematical Physics}\ }\textbf
  {\bibinfo {volume} {158}},\ \bibinfo {pages} {341} (\bibinfo {year}
  {1993})}\BibitemShut {NoStop}%
\bibitem [{\citenamefont {Moessner}\ \emph {et~al.}(2001)\citenamefont
  {Moessner}, \citenamefont {Sondhi},\ and\ \citenamefont
  {Chandra}}]{Moessner01c}%
  \BibitemOpen
  \bibfield  {author} {\bibinfo {author} {\bibfnamefont {R.}~\bibnamefont
  {Moessner}}, \bibinfo {author} {\bibfnamefont {S.~L.}\ \bibnamefont
  {Sondhi}}, \ and\ \bibinfo {author} {\bibfnamefont {P.}~\bibnamefont
  {Chandra}},\ }\href {\doibase 10.1103/PhysRevB.64.144416} {\bibfield
  {journal} {\bibinfo  {journal} {Phys. Rev. B}\ }\textbf {\bibinfo {volume}
  {64}},\ \bibinfo {pages} {144416} (\bibinfo {year} {2001})}\BibitemShut
  {NoStop}%
\bibitem [{\citenamefont {Rokhsar}\ and\ \citenamefont
  {Kivelson}(1988)}]{Rokhsar88}%
  \BibitemOpen
  \bibfield  {author} {\bibinfo {author} {\bibfnamefont {D.~S.}\ \bibnamefont
  {Rokhsar}}\ and\ \bibinfo {author} {\bibfnamefont {S.~A.}\ \bibnamefont
  {Kivelson}},\ }\href@noop {} {\bibfield  {journal} {\bibinfo  {journal}
  {Physical review letters}\ }\textbf {\bibinfo {volume} {61}},\ \bibinfo
  {pages} {2376} (\bibinfo {year} {1988})}\BibitemShut {NoStop}%
\bibitem [{\citenamefont {Woynarovich}\ and\ \citenamefont
  {Eckle}(1987)}]{Woynarovich1987}%
  \BibitemOpen
  \bibfield  {author} {\bibinfo {author} {\bibfnamefont {F.}~\bibnamefont
  {Woynarovich}}\ and\ \bibinfo {author} {\bibfnamefont {H.-P.}\ \bibnamefont
  {Eckle}},\ }\href@noop {} {\bibfield  {journal} {\bibinfo  {journal} {Journal
  of Physics A: Mathematical and General}\ }\textbf {\bibinfo {volume} {20}},\
  \bibinfo {pages} {L97} (\bibinfo {year} {1987})}\BibitemShut {NoStop}%
\bibitem [{\citenamefont {Ralko}\ \emph {et~al.}(2008)\citenamefont {Ralko},
  \citenamefont {Poilblanc},\ and\ \citenamefont {Moessner}}]{Ralko2008}%
  \BibitemOpen
  \bibfield  {author} {\bibinfo {author} {\bibfnamefont {A.}~\bibnamefont
  {Ralko}}, \bibinfo {author} {\bibfnamefont {D.}~\bibnamefont {Poilblanc}}, \
  and\ \bibinfo {author} {\bibfnamefont {R.}~\bibnamefont {Moessner}},\ }\href
  {\doibase 10.1103/PhysRevLett.100.037201} {\bibfield  {journal} {\bibinfo
  {journal} {Phys. Rev. Lett.}\ }\textbf {\bibinfo {volume} {100}},\ \bibinfo
  {pages} {037201} (\bibinfo {year} {2008})}\BibitemShut {NoStop}%
\bibitem [{\citenamefont {Bernu}\ \emph {et~al.}(1992)\citenamefont {Bernu},
  \citenamefont {Lhuillier},\ and\ \citenamefont {Pierre}}]{bernu2}%
  \BibitemOpen
  \bibfield  {author} {\bibinfo {author} {\bibfnamefont {B.}~\bibnamefont
  {Bernu}}, \bibinfo {author} {\bibfnamefont {C.}~\bibnamefont {Lhuillier}}, \
  and\ \bibinfo {author} {\bibfnamefont {L.}~\bibnamefont {Pierre}},\ }\href
  {\doibase 10.1103/PhysRevLett.69.2590} {\bibfield  {journal} {\bibinfo
  {journal} {Phys. Rev. Lett.}\ }\textbf {\bibinfo {volume} {69}},\ \bibinfo
  {pages} {2590} (\bibinfo {year} {1992})}\BibitemShut {NoStop}%
\bibitem [{\citenamefont {Sachdev}(2007)}]{sachdev2007}%
  \BibitemOpen
  \bibfield  {author} {\bibinfo {author} {\bibfnamefont {S.}~\bibnamefont
  {Sachdev}},\ }\href@noop {} {\emph {\bibinfo {title} {Quantum phase
  transitions}}}\ (\bibinfo  {publisher} {Wiley Online Library},\ \bibinfo
  {year} {2007})\BibitemShut {NoStop}%
\bibitem [{\citenamefont {Zamfir}\ \emph {et~al.}(2002)\citenamefont {Zamfir},
  \citenamefont {von Brentano}, \citenamefont {Casten},\ and\ \citenamefont
  {Jolie}}]{Zamphir02}%
  \BibitemOpen
  \bibfield  {author} {\bibinfo {author} {\bibfnamefont {N.~V.}\ \bibnamefont
  {Zamfir}}, \bibinfo {author} {\bibfnamefont {P.}~\bibnamefont {von
  Brentano}}, \bibinfo {author} {\bibfnamefont {R.~F.}\ \bibnamefont {Casten}},
  \ and\ \bibinfo {author} {\bibfnamefont {J.}~\bibnamefont {Jolie}},\ }\href
  {\doibase 10.1103/PhysRevC.66.021304} {\bibfield  {journal} {\bibinfo
  {journal} {Phys. Rev. C}\ }\textbf {\bibinfo {volume} {66}},\ \bibinfo
  {pages} {021304} (\bibinfo {year} {2002})}\BibitemShut {NoStop}%
\bibitem [{\citenamefont {Arias}\ \emph {et~al.}(2003)\citenamefont {Arias},
  \citenamefont {Dukelsky},\ and\ \citenamefont {Garcia-Ramos}}]{Arias03}%
  \BibitemOpen
  \bibfield  {author} {\bibinfo {author} {\bibfnamefont {J.~M.}\ \bibnamefont
  {Arias}}, \bibinfo {author} {\bibfnamefont {J.}~\bibnamefont {Dukelsky}}, \
  and\ \bibinfo {author} {\bibfnamefont {J.~E.}\ \bibnamefont {Garcia-Ramos}},\
  }\href {\doibase 10.1103/PhysRevLett.91.162502} {\bibfield  {journal}
  {\bibinfo  {journal} {Phys. Rev. Lett.}\ }\textbf {\bibinfo {volume} {91}},\
  \bibinfo {pages} {162502} (\bibinfo {year} {2003})}\BibitemShut {NoStop}%
\bibitem [{\citenamefont {Mendels}\ and\ \citenamefont
  {Bert}(2010)}]{Mendels2010}%
  \BibitemOpen
  \bibfield  {author} {\bibinfo {author} {\bibfnamefont {P.}~\bibnamefont
  {Mendels}}\ and\ \bibinfo {author} {\bibfnamefont {F.}~\bibnamefont {Bert}},\
  }\href@noop {} {\bibfield  {journal} {\bibinfo  {journal} {J. Phys. Soc.
  Jpn.}\ }\textbf {\bibinfo {volume} {79}},\ \bibinfo {pages} {011001}
  (\bibinfo {year} {2010})}\BibitemShut {NoStop}%
\bibitem [{\citenamefont {Han}\ \emph {et~al.}(2012)\citenamefont {Han},
  \citenamefont {Helton}, \citenamefont {Chu}, \citenamefont {Nocera},
  \citenamefont {Rodriguez-Rivera}, \citenamefont {Broholm},\ and\
  \citenamefont {Lee}}]{Han2012}%
  \BibitemOpen
  \bibfield  {author} {\bibinfo {author} {\bibfnamefont {T.-H.}\ \bibnamefont
  {Han}}, \bibinfo {author} {\bibfnamefont {J.~S.}\ \bibnamefont {Helton}},
  \bibinfo {author} {\bibfnamefont {S.}~\bibnamefont {Chu}}, \bibinfo {author}
  {\bibfnamefont {D.~G.}\ \bibnamefont {Nocera}}, \bibinfo {author}
  {\bibfnamefont {J.~A.}\ \bibnamefont {Rodriguez-Rivera}}, \bibinfo {author}
  {\bibfnamefont {C.}~\bibnamefont {Broholm}}, \ and\ \bibinfo {author}
  {\bibfnamefont {Y.~S.}\ \bibnamefont {Lee}},\ }\href@noop {} {\bibfield
  {journal} {\bibinfo  {journal} {Nature}\ }\textbf {\bibinfo {volume} {492}},\
  \bibinfo {pages} {406} (\bibinfo {year} {2012})}\BibitemShut {NoStop}%
\bibitem [{\citenamefont {Morita}\ \emph {et~al.}(2008)\citenamefont {Morita},
  \citenamefont {Yano}, \citenamefont {Ono}, \citenamefont {Tanaka},
  \citenamefont {Fujii}, \citenamefont {Uekusa}, \citenamefont {Narumi},\ and\
  \citenamefont {Kindo}}]{Morita2008}%
  \BibitemOpen
  \bibfield  {author} {\bibinfo {author} {\bibfnamefont {K.}~\bibnamefont
  {Morita}}, \bibinfo {author} {\bibfnamefont {M.}~\bibnamefont {Yano}},
  \bibinfo {author} {\bibfnamefont {T.}~\bibnamefont {Ono}}, \bibinfo {author}
  {\bibfnamefont {H.}~\bibnamefont {Tanaka}}, \bibinfo {author} {\bibfnamefont
  {K.}~\bibnamefont {Fujii}}, \bibinfo {author} {\bibfnamefont
  {H.}~\bibnamefont {Uekusa}}, \bibinfo {author} {\bibfnamefont
  {Y.}~\bibnamefont {Narumi}}, \ and\ \bibinfo {author} {\bibfnamefont
  {K.}~\bibnamefont {Kindo}},\ }\href@noop {} {\bibfield  {journal} {\bibinfo
  {journal} {J. Phys. Soc. Jpn.}\ }\textbf {\bibinfo {volume} {77}},\ \bibinfo
  {pages} {043707} (\bibinfo {year} {2008})}\BibitemShut {NoStop}%
\bibitem [{\citenamefont {Ferhat}\ and\ \citenamefont
  {Ralko}(2014)}]{Ralko2013}%
  \BibitemOpen
  \bibfield  {author} {\bibinfo {author} {\bibfnamefont {K.}~\bibnamefont
  {Ferhat}}\ and\ \bibinfo {author} {\bibfnamefont {A.}~\bibnamefont {Ralko}},\
  }\href {\doibase 10.1103/PhysRevB.89.155141} {\bibfield  {journal} {\bibinfo
  {journal} {Phys. Rev. B}\ }\textbf {\bibinfo {volume} {89}},\ \bibinfo
  {pages} {155141} (\bibinfo {year} {2014})}\BibitemShut {NoStop}%
\end{thebibliography}%

\end{document}